\documentstyle[preprint,12pt,aps]{revtex}
\begin{document}
\preprint{MC/TH 96/11}
\title{Effective chiral Lagrangians for spin-1 mesons}
\author{Michael C. Birse}
\address{Theoretical Physics Group, Department of Physics and Astronomy,\\
University of Manchester, Manchester, M13 9PL, U.K.\\}
\maketitle
\vskip 20pt
\begin{abstract}
The commonly used types of effective theory for vector mesons are reviewed and
their relationships clarified. They are shown to correspond to different
choices of field for spin-1 particles and the rules for transforming between
them are described. The importance of respecting chiral symmetry is stressed.
The choice of fields that transform homogeneously under the nonlinear
realisation of chiral symmetry imposes no preconceptions about the types of
coupling for the mesons. This representation thus provides a convenient
framework for relating different theories. It is also used to elucidate the
nature of the assumptions in specific hidden-gauge and massive Yang-Mills
models that have been widely used.
\end{abstract}

\section{Introduction}

At very low energies strong interactions among pions can be described by an
effective Lagrangian based on a chirally symmetric sigma model\cite{dgh}. To
extend such a description to higher energies heavier mesons need to be
incorporated, most notably vector mesons. Various schemes for doing so have
been proposed, differing in the transformation properties of their vector
fields under chiral symmetry. 

Many of these approaches are motivated by the phenomenologically successful
ideas of vector-meson dominance and universal coupling\cite{vmd,current}.
These lead to kinetic terms and couplings for the spin-1 mesons that have the
same forms as in a gauge theory, reflecting the assumed universal coupling of
these mesons to conserved currents. Examples include the ``massive
Yang-Mills"\cite{gg,syracuse,meissner} and ``hidden-gauge" theories\cite{bky}.
In these approaches, low-energy theorems of chiral symmetry place important
constraints on the gauge-type coupling of the $\rho$ meson to two pions. It is
essential that such Lagrangians respect chiral symmetry, otherwise they can
lead to unrealistic results.

However it is not necessary to impose a gauge structure on the effective
Lagrangian from the start. An alternative scheme for incorporating these mesons
was suggested by Weinberg\cite{wein} and developed further by Callan, Coleman,
Wess and Zumino\cite{ccwz}. In this treatment, denoted here by WCCWZ, the
fields transform homogeneously under a  nonlinear realisation of chiral
symmetry. Another related approach is that of Ecker {\it et al.}\ in which the
spin-1 mesons are represented by antisymmetric tensor, rather than vector, 
fields\cite{ecker1,ecker2}. In contrast to the gauge-type theories, these 
formalisms have $\rho\pi\pi$ couplings that involve higher powers of
momentum and are not directly constrained by chiral symmetry.

Despite the rather different forms of their Lagrangians, and the different
types of coupling contained in them, all of these approaches are in principle
equivalent. Each corresponds to a different choice of fields for
the spin-1 mesons. This is illustrated rather well in extended
Nambu--Jona-Lasinio models\cite{enjl,bijnens}, where there is considerable
freedom in the choice of auxiliary fields in the vector and axial channels. To
some extent the choice of scheme must be based on the simplicity of the
resulting Lagrangian.  In making comparisons between the approaches it is
important not to confuse features that arise from the choice of 
field with those that arise from requiring, for instance, universal coupling of
the vector mesons. The former are not physical, controlling merely the
off-shell behaviour of scattering amplitudes. The latter do have physical
consequences, such as relations between on-shell amplitudes for different
processes. 

Effective Lagrangians of spin-1 mesons were extensively reviewed by Meissner
in 1988\cite{meissner}. Since then there have been a number of developments in
the field, most notably in connection with chiral perturbation
theory\cite{ecker1,ecker2}. Moreover interest in these Lagrangians has
recently been reawakened by the possibility that experiments at
high-luminosity accelerators such as CEBAF or DA$\Phi$NE may be able to
explore some of the couplings that have up to now been
inaccessible\cite{bgp,esk}. Another source of interest arises from the use of
these theories in calculations of charge-symmetry violation in nuclear
forces\cite{oconn}. In addition, they are being used in studies of the
behaviour of vector mesons in hot and dense matter\cite{gklsy,sh,br} and some
of them lead to quite different predictions for the mass of the $\rho$ in
matter\cite{pisarski}. In all of these contexts it is important to be able to
compare theories, even though they may be expressed in different formalisms,
in a way that is independent of the different choices of fields. To this end I
explore here the connections between the various approaches, and their
corresponding fields.

In Sec.~II, I review some basic ideas of chiral symmetry, focussing on the 
constraints that the symmetry imposes on interactions with a vector character
through the Weinberg-Tomozawa low-energy theorem\cite{wpin,tomo} for pion
scattering from a target with nonzero isospin and the KSFR relation\cite{ksfr}
for the couplings of the $\rho$ meson. The ingredients needed for construction
of effective Lagrangians using the nonlinear realisation of chiral symmetry are
also outlined.

The WCCWZ scheme\cite{ccwz} is introduced in Sec.~III. I use it throughout
this paper as an overall framework to compare and relate the other approaches
since it imposes no prejudices about the forms of the couplings among the
mesons. As noted by Ecker {\it et al.}\cite{ecker2}, the consequences of
physical assumptions like vector dominance can then be rather transparently
expressed as relations between the couplings in a WCCWZ Lagrangian. By
converting commonly used hidden-gauge and massive Yang-Mills theories into
their WCCWZ equivalents, their couplings can be directly compared.

Within this scheme, the leading contributions to low-energy $\pi\pi$
scattering arise from four-pion interaction terms in the Lagrangian;
$\rho$-exchange contributions are suppressed by powers of the pion momenta.
The coupling constants for these four-pion interactions can thus be determined
from ChPT at order $p^4$\cite{gl,ecker2}. It turns out that their values are
in good agreement with those obtained using the assumption of resonance 
saturation\cite{gl,ecker1,ecker2,drv} in the corresponding channel of $\pi\pi$
scattering. (This assumption is related to, but not as strong as, that of
vector dominance, as discussed in Sec.~III.) Moreover these four-point
interactions are essential if the effective theory is to be well-behaved at
short distances. For example if the Hamiltonian is to be bounded from below,
the four-point couplings should satisfy inequalities relating them to
three-point ones\cite{kal,kb}. The corresponding equalities are then obtained
from the stronger assumption of resonance saturation by a single
meson\cite{ecker2}.

In the hidden-gauge approach\cite{bky}, described in Sec.~IV, an artificial
local symmetry is introduced into the nonlinear sigma model by the choice of
field variables. The $\rho$ meson is then introduced as a gauge boson for this
symmetry. As stressed by Georgi\cite{georgi}, the additional local symmetry
has no physics associated with it, and it can be removed by fixing the gauge.
In the unitary gauge the symmetry reduces to a nonlinear realisation of chiral
symmetry, under which the vector fields transform inhomogeneously, in contrast
to those of WCCWZ. However, with a further change of variable any vector-meson
Lagrangian of the hidden-gauge form can be converted into an equivalent WCCWZ
one\cite{georgi}. The rules for transforming a Lagrangian from hidden-gauge to
WCCWZ form have also been noted by Ecker {\it et al.}\cite{ecker2}.

As I show here, by changing variables from the hidden-gauge to WCCWZ scheme,
the gauge coupling constant of the former is really a parameter in the choice
of vector field. This coupling constant does not appear in the equivalent
WCCWZ Lagrangian and so hidden-gauge theories with different gauge couplings,
together with different higher-order couplings, can be equivalent. The
conventional choice is shown to be one that eliminates any ${\cal O}(p^3)$
$\rho\pi\pi$ coupling from the hidden-gauge Lagrangian, so that the leading
corrections to the ${\cal O} (p)$ coupling are of order $p^5$. If the
$\gamma\rho$ mixing strengths satisfy a particular relation\cite{ecker2}, then
this choice of field also eliminates the leading momentum-dependent
corrections, of order $p^2$, to the mixing. This reduction of the momentum
dependence of the couplings thus allows the hidden-gauge approach to embody
the empirical observation that the KSFR relation\cite{ksfr} is well satisfied
by the $\rho\pi\pi$ and $\gamma\rho$ couplings determined from the decay of
on-shell $\rho$ mesons.

In massive Yang-Mills theories\cite{gg,meissner}, described in Sec.~V, the
vector and axial fields transform under a linear realisation of chiral
symmetry. Three- and four-point couplings among these fields are included and,
together with the kinetic terms, form a Yang-Mills Lagrangian with a local
chiral symmetry. The full theory does not possess this gauge symmetry since it
includes mass terms which have only global symmetry. By changing variables to
spin-1 fields that transform under the nonlinear realisation of chiral
symmetry, any massive Yang-Mills theory can be converted into an equivalent
WCCWZ one and its relations to other theories, such as hidden-gauge ones, can 
be explored. 

The use of a linear realisation of chiral symmetry means that both the $\rho$ 
meson and its chiral partner the $a_1$ must be treated on the same footing.
One cannot simply omit the $a_1$ from a massive Yang-Mills theory without
violating chiral symmetry. Nonetheless it is possible to write down
Lagrangians with a Yang-Mills form for the $\rho$ meson alone, provided that
one takes care to include additional terms that ensure satisfaction of the
chiral low-energy theorems\cite{bpr}. As described here a convenient way to
generate these terms is to take a hidden-gauge theory and make a change of
variables that brings it into a Yang-Mills-like form.

The final formalism I consider is the one based on antisymmetric tensor 
fields\cite{ecker1,ecker2,pp} described in Sec.~VI. These fields transform
homogeneously under the nonlinear realisation of chiral symmetry and so the
approach has many similarities with that of WCCWZ. The main difference is that
the basic $\rho\pi\pi$ and $\gamma\rho$ couplings involve one less power of
momentum. This means that, if resonance saturation is assumed, the Lagrangian
can take a particularly simple form. For every coupling in the general WCCWZ
Lagrangian, one can construct a corresponding one involving tensor
fields\cite{ecker2,bm} A more direct way using path integrals to translate
between the two schemes has been described in\cite{bp}.

In Sec.~VII, I discuss briefly the explicit symmetry-breaking terms that can
appear in these Lagrangians and comment on their applications to 
isospin-violating processes.

\section{Chiral symmetry}

\subsection{PCAC}

The current masses of up and down quarks are very much smaller than
typical hadron energy scales, and hence to a good approximation QCD is 
invariant under both ordinary isospin rotations, 
$$\psi\rightarrow\left(1-{\textstyle{1\over 2}} i\hbox{\boldmath$\beta$}\cdot
\hbox{\boldmath$\tau$}\right)\psi, 
\eqno(2.1)$$
and axial isospin rotations, 
$$\psi\rightarrow\left(1-{\textstyle{1\over 2}} i\hbox{\boldmath$\alpha$}\cdot
\hbox{\boldmath$\tau$}\gamma_5\right)\psi,\eqno(2.2)$$
where {\boldmath$\alpha$} and {\boldmath$\beta$} denote infinitesimal 
parameters. Together these form the chiral symmetry group
SU(2)$_R\times$SU(2)$_L$\cite{dgh}. (I concentrate here on the up and down
quarks; the extension to three light flavours is straightforward.) The
corresponding Noether currents are the (vector) isospin currents $${\bf
J}^\mu=\overline\psi\gamma^\mu{\textstyle{1\over 2}}
\hbox{\boldmath$\tau$}\psi,  \eqno(2.3)$$
and the axial currents
$${\bf J}_5^\mu=\overline\psi\gamma^\mu\gamma_5{\textstyle{1\over 2}}
\hbox{\boldmath$\tau$}\psi. \eqno(2.4)$$
The presence of small current masses for the up and down quarks means that this 
symmetry is only approximate. The axial currents thus have divergences 
proportional to these masses. Furthermore the difference between the up and
down quark masses breaks isospin symmetry and so the vector currents have
nonzero divergences.

Despite the smallness of the current masses, the QCD vacuum is not even
approximately invariant under axial isospin rotations. The chiral symmetry is
hidden (or ``spontaneously broken") and so degenerate states of opposite
parity do not appear in the hadron spectrum. Instead the pions are close to
being the corresponding massless Goldstone bosons. The hidden symmetry shows
up in the forms of the interactions of low-energy pions.

One important consequence of this hidden symmetry is the nonzero matrix
element for the weak decay of charged pions
$$\langle 0|J_5^{a\mu}(x)|\pi_b(q)\rangle= i f_\pi q^\mu e^{- i q\cdot x} 
\delta_{ab}, \eqno(2.5)$$
where the pion decay constant is $f_\pi=92.4\pm 0.3$ MeV \cite{pdg}. 
The divergence of this equation is 
$$\langle 0|\partial_\mu J_5^{a\mu}(x)|\pi_b(q)\rangle=f_\pi m_\pi^2 
e^{- i q\cdot x} \delta_{ab}, \eqno(2.6)$$
which shows that the operators 
$$\hbox{\boldmath$\phi$}(x)=\partial_\mu{\bf J}_5^\mu(x)/(f_\pi m_\pi^2) 
\eqno(2.7)$$
connect the vacuum and one-pion states with the same normalisation that
canonical pion fields would have. These interpolating fields provide the
basis for an approach known as ``partial conservation of the axial current"
(PCAC). This is a method for elucidating the consequences of approximate chiral
symmetry for the interactions of low-energy pions.

Of course the use of these particular interpolating fields is a matter of
choice. Other pion fields should give the same results for all physical
amplitudes involving on-shell pions; where they differ is in their off-shell
extrapolations. The advantage of the PCAC choice is that, in the soft-pion
limit, we can relate amplitudes for interactions with pions to the axial
transformation properties of the states involved.

For the present discussion of vector mesons, it is worth re-examining two of
the consequences of PCAC. First consider the scattering of pions off a
target with nonzero isospin. By applying LSZ reduction to the amplitude for
forward scattering of a pion with momentum $k$ off a target with momentum $p$,
it can be written in the form (see, for example,\cite{bc})
$$F^{ab}=i\left({m_\pi^2-k^2\over f_\pi m_\pi^2}\right)^2\int d^4x\, 
e^{ik\cdot x}\langle p|T\Bigl(\partial_\mu J_5^{a\mu}(x),\partial_\nu 
J_5^{b\nu}(x)\Bigr)|p\rangle, \eqno(2.8)$$
where I have used the PCAC pion fields defined by (2.7). On integrating by
parts, the integral gives rise to two terms. One of these is an equal-time
commutator which reduces to an explicit chiral symmetry-breaking matrix
element (or ``sigma commutator") in the soft-pion limit\cite{wpin}. This piece
is of order $m_\pi^2$. The second term, which has a factor of $k_\mu$, can be
integrated by parts again to give two terms: another equal time commutator and
term with a factor of $k_\mu k_\nu$. The first of these contains a piece that
is linear in the pion energy and so forms the leading term in a chiral
expansion of the amplitude.

This leading term can be found by taking the soft-pion limit, that is setting
the pion three-momentum to zero and letting the energy tend to zero. The
amplitude (2.8) (integrated by parts as just described) then reduces to
$$F^{ab}={1\over f_\pi^2}k^0\langle p|[J_5^{a0}(0),Q_5^b]|p\rangle
+{\cal O}(m_\pi^2,k^2), \eqno(2.9)$$
where $Q_5^a$ is the axial charge corresponding to (2.4). Using the algebra
associated with the SU(2)$_R\times$SU(2)$_L$ symmetry group, the commutator in 
this expression is just
$$[J_5^{a0}(0),Q_5^b]=i\epsilon^{abc}J^{c0}(0). \eqno(2.10)$$
Hence the leading term in the scattering amplitude has the form
$$F^{ab}={i\over f_\pi^2}k^0\epsilon^{abc}\langle p|J^{c0}|p\rangle
+{\cal O}(m_\pi^2,k^2). \eqno(2.11)$$
Since $-i\epsilon^{abc}$ are just the matrix elements of the pion isospin 
operator, we can see that this amplitude is proportional to the scalar product
of the isospin operators for the pion and target. This is the famous result
derived by Weinberg and Tomozawa for low-energy pion-nucleon
scattering\cite{wpin,tomo}. The argument here shows that this form is general
and should be present for pion scattering from any target.

The form of the Weinberg-Tomozawa term is exactly the same as one would get
from exchange of a $\rho$ meson coupled to the isospin currents of the pion
and target. However it is important to remember that it arises as a consequence
of chiral symmetry and has no necessary connection with $\rho$ exchange.
Moreover its strength is fixed by the symmetry alone. Hence in any approach
where $\rho$ meson is coupled to the isospin currents, as in a gauge theory,
the requirement that $\rho$ exchange does not violate this low-energy theorem
places constraints on the couplings of the $\rho$ to other particles.

A related result can be obtained from the amplitude for $\rho$ decay into
two pions. Applying an LSZ reduction to this as above, it can be written as
$$G^{ab}=i\left({m_\pi^2-k^2\over f_\pi m_\pi^2}\right)^2\int d^4x\, 
e^{ik\cdot x}\langle 0|T\Bigl(\partial_\mu J_5^{a\mu}(x),\partial_\nu 
J_5^{b\nu}(x)\Bigr)|\rho(p)\rangle. \eqno(2.12)$$
In the real world where the $\rho$ is massive, there is no way to
extrapolate this amplitude to the soft-pion limit of vanishing four-momenta.
Nonetheless, at least in the context of an effective theory where the mass of
the $\rho$ appears as a parameter in the Lagrangian, one can ask how this
amplitude should behave as the $\rho$ mass is taken by hand to zero.

After manipulating this amplitude in the same way as for the pion scattering
above, it can be written in the form
$$G^{ab}={i\over f_\pi^2}k^0\epsilon^{abc}\langle 0|J^{c0}|\rho(p)\rangle
+{\cal O}(m_\pi^2k,k^3), \eqno(2.13)$$
in this artificial soft-pion, light-$\rho$ limit. This relation links the
$\rho\pi\pi$ coupling at first order in $k$ to the $\rho$-to-vacuum matrix
element of the vector current. The vector, isovector nature of the $\rho$
means that there is no term analogous to the sigma commutator and so
higher-order terms start at third order in $k$ and $m_\pi$. The matrix element
in (2.13) is responsible for the electromagnetic decay $\rho^0\rightarrow
e^+e^-$ and can be expressed in the form
$$\langle 0|J^{c0}|\rho(p)\rangle=g_{\rho\gamma}(m_\rho). \eqno(2.14)$$ 

In the soft-pion limit, the only contribution to $\rho\rightarrow\pi\pi$ of
order $k$ arises from the coupling of the $\rho$ to the pionic isospin current
$${\cal L}_{\rho\pi\pi}=g_{\rho\pi\pi}\hbox{\boldmath$\rho$}^\mu\cdot
\hbox{\boldmath$\pi$}\wedge\partial_\mu\hbox{\boldmath$\pi$}, \eqno(2.15)$$
if such a term is present in the effective Lagrangian. Comparing the 
corresponding amplitude $G^{ab}$ with (2.13) gives
$$g_{\rho\gamma}(0)=2g_{\rho\pi\pi}f_\pi^2, \eqno(2.16)$$
where both $\rho$ couplings are evaluated in the soft-pion, light-$\rho$
limit. This is the celebrated KSFR relation\cite{ksfr} in its first form. For
any chirally-symmetric effective Lagrangian it relates the strength of
$\gamma\rho$ mixing at zero four-momentum to that of the coupling in (2.15).

Although the KSFR relation has been derived by taking an unphysical limit
where the $\rho$ mass tends to zero, one can ask how close the real world is
to this limit. Rather remarkably in view of the large $\rho$ mass, if one
assumes that the $\gamma\rho$ mixing is independent of momentum and the
$\rho\pi\pi$ coupling is purely of the form (2.15), then the coupling
strengths deduced from the observed $\rho$ decays satisfy (2.16) to about
10\%. This suggests that it should be possible to find a realistic effective
Lagrangian for $\pi\rho$ physics in which higher-order corrections to these
couplings are small. As we shall see in Sec.\ IV, the hidden-gauge scheme
can provide just such a Lagrangian.

If one further assumes complete vector dominance in the photon-pion
coupling\cite{vmd,current}, as described in more detail in Sec.~III, then the
$\rho$ couplings also satisfy
$$g_{\rho\gamma}={m_\rho^2\over g_{\rho\pi\pi}}. \eqno(2.17)$$
This can be combined with (2.16) to obtain a second form of the KSFR 
relation,
$$m_\rho^2=2g_{\rho\pi\pi}^2f_\pi^2, \eqno(2.18)$$
which is also satisfied remarkably well by the coupling deduced from $\rho$
decay. If one assumes vector dominance in the couplings of the photon to all
hadrons, which demands a universal coupling of the $\rho$ to the isospin
current, then this relation has a further interesting consequence: $\rho$
exchange between a pion and another hadron, such as a nucleon, can account for
the whole strength of the Weinberg-Tomozawa term discussed above. 

\subsection{Effective Lagrangians}

The PCAC pion field is useful in studying processes that involve external
pions. However it is not a canonical pion field (except in certain rather
simple models) and so cannot be used to calculate the effects of virtual
pions, either as exchanged particles or in loop diagrams. For these it is most
convenient to work with effective field theories that embody the constraints 
imposed by chiral symmetry on the couplings between the fields.\footnote{A 
modern review of these ideas can be found in the book by Donoghue, Golowich
and Holstein\cite{dgh}.}

The simplest way to construct a Lagrangian satisfying chiral symmetry is to 
introduce fields that transform linearly under the group
SU(2)$_R\times$SU(2)$_L$, in the same way that the corresponding quark
bilinears do. In particular, pion fields {\boldmath$\phi$}$(x)$ can be
introduced along with a scalar field $\sigma(x)$ to form a multiplet
$(\sigma,\hbox{\boldmath$\phi$})$ that transforms like
$(\overline\psi\psi,\overline\psi i\gamma_5 \hbox{\boldmath$\tau$}\psi)$.
The excitations of these fields are pions, which are almost massless,
approximate Goldstone bosons, together with massive scalar mesons. If we are 
only interested in low-energy physics then we may eliminate the massive
degrees of freedom by restricting these fields to the chiral circle,
$\sigma(x)^2+\hbox{\boldmath$\phi$}(x)^2=f_\pi^2$.

The Lagrangian for the resulting nonlinear sigma model can be expressed in
terms of the SU(2) matrix $U(x)$ defined by $f_\pi U(x)=\sigma(x)
+i\hbox{\boldmath$\tau\cdot\phi$(x)}$. Under the global SU(2)$\times$SU(2)
chiral symmetry this transforms as
$$U(x)\to g_L U(x) g_R^\dagger, \eqno(2.19)$$
with $g_L,g_R \in SU(2)$. The matrix field $U(x)$ can be parametrised in terms
of three pion fields in a variety of ways. Here I shall use the exponential
form $U(x)=\exp(i\hbox{\boldmath$\tau$}\cdot\hbox{\boldmath$\pi$} (x)/f_\pi)$.

In the chiral limit, the leading piece of the Lagrangian is
$${\cal L}={f_\pi^2\over 4}\langle\partial_\mu U\partial^\mu U\rangle, 
\eqno(2.20)$$
where $\langle \cdots\rangle $ denotes a trace in SU(2) space. This should be
supplemented with terms describing explicit chiral symmetry breaking and
interactions of higher-order in the pion momenta\cite{gl,dgh}. Estimates of
the coefficients of these interactions based on the idea of resonance
saturation\cite{gl,ecker1,drv} agree well with phenomenologically determined
values. This has led people to consider extended Lagrangians that include
fields describing the heavier particles whose exchanges can generate the
higher-order interactions. Amongst these Lagrangians are the ones for spin-1
mesons discussed here.

The most straightforward way to introduce the spin-1 $\rho$ and $a_1$ mesons 
is to use fields $\widetilde{\hbox{\bf V}}^\mu$ and $\widetilde{\hbox{\bf 
A}}^\mu$ that transform like the vector and axial currents
(2.3,4).\footnote{These fields cannot however be simply identified with the
corresponding currents since, like the PCAC interpolating pion fields, these
currents are not canonical field operators.} The transformation properties of
these are most easily given in terms of their right- and left-handed
combinations
$$\widetilde{\hbox{\bf X}}^\mu={1\over\sqrt 2}\left(\widetilde{\hbox{\bf 
V}}^\mu+\widetilde{\hbox{\bf A}}^\mu\right),\qquad 
\widetilde{\hbox{\bf Y}}^\mu={1\over\sqrt 2}\left(\widetilde{\hbox{\bf V}}^\mu
-\widetilde{\hbox{\bf A}}^\mu\right). \eqno(2.21)$$
Under the chiral rotation above, these fields become
$$\widetilde X_\mu\rightarrow g_R\widetilde X_\mu g_R^\dagger$$
$$\widetilde Y_\mu\rightarrow g_L\widetilde Y_\mu g_L^\dagger, \eqno(2.22)$$
where I have introduced matrix fields, $X_\mu={1\over
2}\hbox{\boldmath$\tau$}\cdot\hbox{\bf X}_\mu$, and so on.
Such linearly transforming fields are the basis for the massive Yang-Mills
theories described in Sec.~V.

\subsection{Nonlinear realisation}

Although the transformation properties of the fields in (2.22) are simple,
the couplings of these fields to pions do not necessarily vanish in the
soft-pion limit. As a result, calculations of scattering amplitudes involve
large contributions with strong cancellations between them which are needed to
satisfy chiral low-energy theorems. For many purposes it would be simpler if
the correct low-energy behaviour was already built into the interaction terms
appearing in the Lagrangian. This can be achieved by switching to the
nonlinear realisation of chiral symmetry introduced by Weinberg\cite{wein},
which forms the basis for both the WCCWZ and hidden-gauge schemes. This
realisation of the symmetry is obtained from the transformation properties of
the square root of $U(x)$, denoted by $u(x)$:
$$u \to g_L u\, h^\dagger\left(u,g_L,g_R\right)=h\left(u,g_L,g_R
\right)u g_R^\dagger, \eqno(2.23)$$
where $h\left(u(x),g_L,g_R\right)$ is a compensating SU(2) rotation which
depends on the pion fields at $x$ as well as on $g_{L,R}$. The detailed form
of $h$ is not needed here; it can be found in Ref.\cite{ccwz}. 

It is convenient to introduce the following field gradients 
$$u_\mu=i (u^\dagger\partial_\mu u-u\partial_\mu u^\dagger)$$
$$\Gamma_\mu={1\over 2}(u^\dagger\partial_\mu u+u\partial_\mu u^\dagger).
\eqno(2.24)$$
If these are expanded in powers of the pion field {\boldmath $\pi$}$(x)$, we
find that the leading terms are
$$u_\mu=-{1\over f_\pi}\hbox{\boldmath$\tau$}\cdot\partial_\mu
\hbox{\boldmath$\pi$}+\cdots,$$
$$\Gamma_\mu={i\over 4f_\pi^2}\hbox{\boldmath$\tau\cdot\pi$}\wedge\partial_\mu
\hbox{\boldmath$\pi$}+\cdots, \eqno(2.25)$$
from which we can see that they are an axial vector and vector respectively. 
Under the nonlinear realisation of chiral symmetry the transformation of 
$u_\mu$ is homogeneous 
$$u_\mu \to h u_\mu h^\dagger, \eqno(2.26)$$
whereas that of $\Gamma_\mu$ is inhomogeneous,
$$\Gamma_\mu \to h\Gamma_\mu h^\dagger+h\partial_\mu h^\dagger.\eqno(2.27)$$
The simple sigma model (2.20) can be expressed in terms of $u^\mu$ as
$${\cal L}={f_\pi^2\over 4}\langle u_\mu u^\mu\rangle, \eqno(2.28)$$
which from (2.26) is chirally invariant.

In fact $\Gamma_\mu$ is the connection on the coset space 
SU(2)$\times$SU(2)/SU(2) and and with it we can construct the covariant
derivative on this space:
$$\nabla_{\mu}=\partial_{\mu} + [\Gamma_\mu,\ ].\eqno(2.29)$$ 
The covariant derivatives of $u_\mu$ satisfy the useful relation
$$\nabla_\mu u_\nu-\nabla_\nu u_\mu=0. \eqno(2.30)$$
Also, the curvature tensor corresponding to $\Gamma_\mu$ can be expressed in
terms of $u_\mu$ as
$$\partial_\mu\Gamma_\nu-\partial_\nu\Gamma_\mu+[\Gamma_\mu,\Gamma_\nu]=
{1\over 4}[u_\mu,u_\nu]. \eqno(2.31)$$

Vector and axial fields can be defined that transform homogeneously under this
realisation symmetry, like $u^\mu$,
$$V_\mu \to h V_\mu h^\dagger$$
$$ A_\mu \to h A_\mu h^\dagger. \eqno(2.32)$$ 
Such fields can be obtained from the linearly transforming ones described 
in (2.21) above by multiplying them by $u(x)$ and its conjugate:
$$X_\mu=u\widetilde X_\mu u^\dagger$$
$$Y_\mu=u^\dagger\widetilde Y_\mu u. \eqno(2.33)$$
These form the basis of the WCCWZ scheme described in the next section.

\subsection{External fields}

To describe electromagnetic and weak interactions, we also need to couple our
hadron fields to external vector and axial fields. In the case of linearly 
realised chiral symmetry this is straightforward. The minimal couplings are
obtained by replacing all derivatives with the corresponding gauge-covariant
derivatives, for example:
$$\partial_\mu U\rightarrow D_\mu U=\partial_\mu U+i[U,v_\mu]
+i[U,a_\mu]_+,\eqno(2.34)$$
where $v_\mu$ and $a_\mu$ are the external fields. (Factors of coupling
constants such as $e$ have been absorbed into the definitions of these fields.)

Other, nonminimal couplings, such as anomalous magnetic moments for the vector
mesons, can be included by adding terms to the Lagrangian that involve the
gauge-invariant field strengths $v_{\mu\nu}$ and $a_{\mu\nu}$. There should
also be direct $\gamma\rho$ mixing. This can be included in a 
gauge-invariant manner through a term of the form
$${\cal L}_{\rho\gamma(k)}=-{g_{\rho\gamma}\over m_\rho^2}\langle 
V^{\mu\nu}v_{\mu\nu}\rangle, \eqno(2.35)$$
where the value of $g_{\rho\gamma}$ is that for an on-shell $\rho$-meson.
By making appropriate changes of variable such a Lagrangian can also be
rewritten in a form where the mixing arises from a mass-type
term\cite{vmd,current,klz} 
$${\cal L}_{\rho\gamma(m)}=-2g_{\rho\gamma}\langle V^\mu v_\mu\rangle, 
\eqno(2.36)$$ 
where the gauge invariance is no longer transparent. If vector dominance
(2.17) holds exactly, then this transformed Lagrangian contains no direct
$\gamma\pi\pi$ coupling at first order in the pion momentum: the entire photon
coupling to the pion arises from a virtual $\rho$.

The couplings discussed so far are those of isovector external fields. 
Electromagnetic interactions also contain isoscalar pieces, which are rather
different in character. The leading isoscalar coupling of the photon is to
three pions and this can be described by a term coupling the photon to the 
topological current
$$B^\mu={1\over 24\pi^2}\epsilon^{\mu\nu\kappa\lambda}\langle (U^\dagger
\partial_\nu U)(U^\dagger\partial_\kappa U)(U^\dagger\partial_\lambda U)
\rangle, \eqno(2.37)$$
which provides the baryon number current in the Skyrme model\cite{skyrme,gw}.
This coupling forms part of the anomalous Lagrangian, which also describes
processes such as $\pi^0 \rightarrow\gamma\gamma$. Unlike the ones discussed
so far, these terms are odd under the operation $U(x)\rightarrow
U(x)^\dagger$\cite{witten}. In the case of an effective theory with SU(3)
symmetry, the $\gamma \pi^+\pi^0\pi^-$ coupling can be obtained from the
gauged version of the Wess-Zumino-Witten term\cite{wz,witten} (see
also:\cite{zb,meissner}), although for an SU(2) theory it has to be added by
hand.

With the nonlinear realisation of the symmetry, the use of gauge-covariant
derivatives means that the definitions of the gradients (2.24) should be
replaced by
$$u_\mu=iu^\dagger[\partial_\mu-i(v_\mu-a_\mu)]u-iu[\partial_\mu
-i(v_\mu+a_\mu)]u^\dagger)$$
$$\Gamma_\mu={1\over 2}\left(u^\dagger[\partial_\mu-i(v_\mu-a_\mu)]u
+u[\partial_\mu-i(v_\mu+a_\mu)]u^\dagger\right). \eqno(2.38)$$
In the nonminimal couplings, the field-strength tensors should appear in the
nonlinearly transforming combinations,
$$F_\pm^{\mu\nu}={1\over 2}\left(u(v_{\mu\nu}+a_{\mu\nu})u^\dagger
\pm u^\dagger(v_{\mu\nu}-a_{\mu\nu})u\right), \eqno(2.39)$$
where $F_+^{\mu\nu}$ is the combination that corresponds to a vector field
coupling and $F_-^{\mu\nu}$ to an axial one. The tensor $F_+^{\mu\nu}$ also
appears as an additional term in the curvature tensor defined using the
covariant derivatives on the coset space:
$$\partial_\mu\Gamma_\nu-\partial_\nu\Gamma_\mu+[\Gamma_\mu,\Gamma_\nu]=
{1\over 4}[u_\mu,u_\nu]-iF_{+\mu\nu}.\eqno(2.40)$$

Explicit symmetry breaking arising from the current quark masses can be
introduced by treating those masses as if they were uniform external scalar
fields. In the nonlinear realisation, the corresponding terms in the 
Lagrangian can be expressed in terms of the quantities
$$\chi_\pm=u{\cal M}u\pm u^\dagger{\cal M}u^\dagger, \eqno(2.41)$$
where ${\cal M}$ is proportional to the matrix of current quark masses.

\section{WCCWZ}

In the WCCWZ approach\cite{wein,ccwz,ecker2} vector and axial fields transform
homogeneously under the nonlinear realisation of chiral symmetry just
described. This scheme imposes no constraints on the couplings of the spin-1
mesons, apart from those that follow from approximate chiral symmetry. 

Denoting the fields describing the $\rho$ and $a_1$ mesons by $V^\mu$ and 
$A^\mu$ and defining their covariant derivatives by
$$V_{\mu\nu}=\nabla_\mu V_\nu-\nabla_\nu V_\mu,\qquad 
A_{\mu\nu}=\nabla_\mu A_\nu-\nabla_\nu A_\mu, \eqno(3.1)$$  
we can then express the kinetic and mass terms of the Lagrangian as
$${\cal L}=-{1\over 2}\langle V_{\mu\nu}V^{\mu\nu}\rangle 
-{1\over 2}\langle A_{\mu\nu}A^{\mu\nu}\rangle 
+m_V^2\langle V_\mu V^\mu\rangle+m_A^2\langle A_\mu A^\mu\rangle. \eqno(3.2)$$
Expanding the covariant derivatives here to leading order in the pion fields,
we find that this contains the term
$$-4\langle\partial_\mu V_\nu[\Gamma^\mu, V^\nu]\rangle
={1\over f_\pi^2}(\hbox{\bf V}^\nu\wedge\partial_\mu\hbox{\bf V}_\nu)\cdot
(\hbox{\boldmath$\pi$}\wedge\partial^\mu\hbox{\boldmath$\pi$}) \eqno(3.3)$$
which corresponds to a local coupling between the isospin currents of the pions
and $\rho$ mesons. In low-energy $\pi\rho$ scattering, this is just what is 
needed to give the Weinberg-Tomozawa term (2.11). Note that in the WCCWZ
scheme this piece is contained within the kinetic term of the Lagrangian and
is generated without $\rho$-exchange contributions. Moreover all other
contributions to $\pi\rho$ scattering arise from vertices with at least two
factors of $u_\mu$ (to satisfy the restrictions of parity and isospin) and so
are suppressed by at least one further power of the pion momentum.\footnote{If
explicit chiral symmetry breaking terms are included then these also
contribute at next-to-leading order in the chiral expansion, since they are
proportional to $m_\pi^2$.}

A general chirally symmetric Lagrangian for $\pi\rho a_1$ physics consists of
all terms that can be constructed out of traces of products of $u_\mu$,
$V_\mu$, $A_\mu$ and their covariant derivatives, and that are symmetric under 
parity. The derivatives can include both the antisymmetric ones of (3.1) and
the symmetric combinations, as recently pointed out by Borasoy and 
Meissner\cite{bm}. Up to fourth-order in pion-field gradients and the vector
fields, the WCCWZ Lagrangian includes the terms
$${\cal L}={f_\pi^2\over 4}\langle u_\mu u^\mu\rangle -{1\over 2}\langle 
V_{\mu\nu}V^{\mu\nu}\rangle +m_V^2\langle V_\mu V^\mu\rangle -{i\over
2} g_1 \langle V_{\mu\nu} [u^\mu,u^\nu]\rangle  +{i\over 2} g_2 \langle
V_{\mu\nu} [V^\mu,V^\nu]\rangle$$
$$ +{1\over 8} g_3 \langle[u_\mu,u_\nu]^2\rangle  -{1\over 4} g_4 \langle
[u_\mu,u_\nu][V^\mu,V^\nu]\rangle  +{1\over 8} g_5 \langle
[V_\mu,V_\nu]^2\rangle +\cdots.\eqno(3.4)$$ 
Even to this order this expression is not complete, but the terms written out
explicitly here are those we shall need below in discussing the connection to
the hidden-gauge Lagrangian of Bando {\it et al.}\cite{bando1}. These include
the famous $\langle[u_\mu,u_\nu]^2\rangle$ term introduced by Skyrme to
stabilise solitons in a nonlinear sigma model\cite{skyrme}. Obviously many
other three- and four-point interactions, involving the axial as well as the
vector field, should also be present in the full effective Lagrangian.

The leading $\rho\pi\pi$ coupling term in this scheme is 
$$-{i\over 2} g_1 \langle V_{\mu\nu} [u^\mu,u^\nu]\rangle =g_1\partial^\mu
\hbox{\bf V}^\nu\cdot\partial_\mu\hbox{\boldmath$\pi$}\wedge\partial_\nu
\hbox{\boldmath$\pi$}, \eqno(3.5)$$
which is of order $p^3$. Hence, as just mentioned, $\rho$ exchange does not
contribute to the Weinberg-Tomozawa term in $\pi\rho$ scattering. It also
means that the first contribution of $\rho$ exchange to low-energy $\pi\pi$
scattering is of order $p^6$, as noted by Ecker {\it et al.}\cite{ecker2}. The
leading contribution to $\pi\pi$ scattering is thus the Skyrme term, which
is of order $p^4$, and the corresponding coupling constant can be taken from
analyses of $\pi\pi$ scattering using ChPT to that order\cite{gl,ecker1,drv}.
The values of this and other four-pion interactions are in good agreement with
those obtained by assuming resonance saturation.

A general Lagrangian in this scheme can suffer from unphysical behaviour at
short distances unless its coupling constants satisfy certain
relations\cite{ecker2,kal,kb}. A particularly transparent way to see this is
to consider the corresponding Hamiltonian, as pointed out by
Kalafatis\cite{kal}. In the presence of certain configurations of mean fields,
the Hamiltonian for high-momentum modes of the vector fields becomes unbounded
from below unless a Skyrme term is present with a coefficient satisfying
$$g_3\geq g_1^2. \eqno(3.6)$$
Analogous inequalities can also be derived that relate other four-point 
couplings to three-point couplings involving axial mesons\cite{kb}.

Although the full construction of the Hamiltonian is somewhat tedious, a quick 
way to obtain these inequalities can be obtained by looking at the terms in
the Lagrangian that involve time-derivatives of the fields. For example, to
obtain (3.6) we need the kinetic energies and fourth and sixth terms of (3.4).
By completing the square we can write these in the form
$${\cal L}={f_\pi^2\over 4}\langle u_\mu u^\mu\rangle -{1\over 2}\left\langle 
\left(V_{\mu\nu}+{i\over 2}g_1[u_\mu,u_\nu]\right)^2\right\rangle
+{1\over 8} (g_3-g_1^2) \langle[u_\mu,u_\nu]^2\rangle +\cdots. \eqno(3.7)$$
The Skyrme term contains a positive-definite piece of second order in both the
time and space derivatives of the pion fields. If the spatial gradients of
these are large enough, this term will dominate over the normal kinetic term
which is also quadratic in the time derivatives. For the overall coefficient of
these is to remain positive for all field configurations, the coefficient of
the Skyrme term must not be negative and hence the couplings must satisfy
(3.6).

This unwanted behaviour can also show up in scattering processes as violations
of unitarity. In an effective theory of $\pi$ and $\rho$ mesons
only, such as (3.4), high-energy $\pi\pi$ scattering can violate the Froissart
bound\cite{fro} (a consequence of unitarity) unless the coefficient of the
Skyrme term satisfies the equality in (3.6)\cite{ecker2}.\footnote{For a more
detailed discussion of dispersion relations for $\pi\pi$ scattering and their
relation to the parameters of ChPT, see\cite{penn}.} This value for $g_3$ is
the one that follows from resonance saturation by the $\rho$
meson\cite{gl,ecker1,ecker2,drv}. The equalities are thus consequences of the
strong assumption of resonance saturation, namely that only one meson
contributes in each of the relevant channels for $\pi\pi$ and other scattering
processes. More generally, if several states with the same quantum numbers
contribute, only the inequalities are satisfied. 

Although it is rather surprising that a low-energy effective theory of this
type should still give reasonable behaviour for high-energy processes, the
values of the couplings that follow from this assumption do agree with those
determined from low-energy scattering processes. The reasons for its success,
like those of vector dominance and KSFR, remain mysterious.

Electromagnetic couplings of the pions and spin-1 mesons can be introduced in
this scheme as described in Sec.~IID. The usual minimal couplings of the
photon to the vector current arise from the use of the field gradients of
(2.38). Nonminimal couplings can include, for example, an anomalous magnetic
moment for the $\rho$, described by a term of the form
$\langle[V_\mu,V_\nu]F_+^{\mu\nu} \rangle$. 

There can also be $\gamma\rho$ mixing, which is of interest in the context of
vector dominance. At lowest-order, this is described by the term 
$${\cal L}_{\rho\gamma}=-f_1\langle V_{\mu\nu}F_+^{\mu\mu}\rangle,
\eqno(3.8)$$
which is chirally and gauge invariant. As well as providing $\gamma\rho$
mixing, this term contributes to the decay $\rho^0\rightarrow\pi^+\pi^-\gamma$,
where its effects can be seen near the endpoint in the photon
spectrum\cite{hn}.

Such a kinetic mixing is of order $p^2$ and so vanishes in the light-$\rho$
limit used to obtain the KSFR relation in Sec.~IIA. As we have
already seen, the leading $\rho\pi\pi$ coupling in this scheme is of order
$p^3$, hence $g_{\gamma\rho}$ and $g_{\rho\pi\pi}$ both vanish as the $\rho$
mass is taken to zero, while all the coupling constants are held fixed. This
confirms that the KSFR relation (2.16) is indeed satisfied in the WCCWZ
approach, although not in a way that has any practical use, because of the
strong momentum dependence of the couplings.

With mixing of the form (3.8), the effective theory can display unphysical
short-distance behaviour, analogous to that discussed above. In this case the
conflict is not with general principles, such as unitarity or existence of a
vacuum, but with QCD predictions for the behaviour of current-current 
correlators at high momentum\cite{fnr79}. For example, the Lagrangian (3.4) 
with mixing (3.8) (as well as minimal couplings) gives a pion electromagnetic
form factor of
$$F_\pi(q^2)=1+{g_1f_1\over f_\pi^2}{q^4\over M_V^2-q^2}. \eqno(3.9)$$
This grows at large $q^2$ as does the corresponding contribution of $\pi\pi$
intermediate states to the current-current correlator. This is inconsistent
with the QCD expectation that it tend to a constant.

The cure is again to include extra terms in the Lagrangian, in this
case\cite{ecker2}
$${\cal L}_{\gamma\pi(nm)}= -{i\over 2} f_2 \langle F_+^{\mu\nu} 
[u_\mu,u_\nu]\rangle-{1\over 2}f_3\langle F_{+\mu\nu}F_+^{\mu\nu}\rangle. 
\eqno(3.10)$$
The second term here is needed to correct the unphysical contribution from
the $\rho$ meson to the current-current correlator. As above, an easy way to
determine the coefficients for these terms is to complete the square, and
write the relevant pieces of the Lagrangian as
$${\cal L}=-{1\over 2}\left\langle \left(V_{\mu\nu}+{i\over 2}g_1[u_\mu,u_\nu]
+f_1F_{+\mu\nu}\right)^2\right\rangle
-{1\over 2} (f_3-f_1^2) \langle F_+^{\mu\nu}F_{+\mu\nu}\rangle $$
$$-{i\over 2}(f_2-g_1f_1)\langle[u_\mu,u_\nu]F_+^{\mu\nu}\rangle+\cdots. 
\eqno(3.11)$$
In this form the momentum-dependent $\gamma\rho$ mixing could be removed by a
change of variables\cite{vmd,current,klz}. However we do not have to perform
this transformation here; all we need to note is that the unwanted
high-momentum behaviour now arises only from the final two terms. It can thus
be eliminated if their coefficients are zero. For example, the corresponding
expression for the pion form factor is
$$F_\pi(q^2)=1+{g_1f_1\over f_\pi^2}{q^4\over M_V^2-q^2}+{f_2\over f_\pi^2}q^2,
\eqno(3.12)$$
from which we can see that $f_2$ should satisfy 
$$f_2=g_1f_1,\eqno(3.13)$$
if the form factor is not to grow at large $q^2$. Similar arguments can
be used to show that  $f_3$ should be given by
$$f_3=f_1^2. \eqno(3.14)$$
Implicit in these relations is the assumption of resonance saturation, since 
$f_2$ and $f_3$ should in principle also contain contributions from other
states with the quantum numbers of the $\rho$. Under similar assumptions about
the axial couplings, involving the $a_1$ meson, there can be another
contribution to $f_3$, of opposite sign\cite{ecker2}.

The stronger assumption of complete vector dominance in the form 
factor\cite{vmd,current} would require that $F_\pi(q^2)$ be given by
$$F_\pi(q^2)={M_V^2\over M_V^2-q^2}. \eqno(3.15)$$
This expression (3.12) can be put into this form if the coupling constants and
$\rho$ mass are related by
$$g_1f_1={f_\pi^2\over M_V^2}, \eqno(3.16)$$
in addition to (3.13). This is the WCCWZ analogue of the relation (2.17) in
models with minimal momentum dependence in the $\rho$ couplings.

The full WCCWZ Lagrangian also contains an anomalous part, whose form I shall
not discuss in detail. This includes terms like the photon coupling to the
topological baryon current (2.37), which can expressed in the form
$$B^\mu={i\over 24\pi^2}\epsilon^{\mu\nu\kappa\lambda}\langle u_\nu u_\kappa 
u_\lambda\rangle. \eqno(3.17)$$
There can also be a coupling of the isoscalar $\omega$ meson to this current,
as well as further anomalous terms involving both $\rho$ and $\omega$ mesons.
Within the WCCWZ framework, there is no danger that such terms can violate
the low-energy theorems for processes such as $\pi^0\rightarrow\gamma\gamma$
or $\gamma\rightarrow3\pi$ because of the ${\cal O}(p^2)$ nature of the 
photon-vector-meson mixing.

\section{Hidden-gauge theories}

In the simplest version of the hidden-gauge approach, the gauge symmetry is
just SU(2) and only vector mesons are treated as gauge bosons\cite{bando1}.
This symmetry is introduced by factorising $U(x)$ into two SU(2) 
matrices\cite{bando2,meissner},
$$U(x)=\xi_L(x)^\dagger \xi_R(x). \eqno(4.1)$$
Since at each point in space-time this factorisation is arbitrary, these
fields are invariant under local SU(2) rotations. A gauge symmetry has thus
been created by this choice of variables.

The extension to axial-vector mesons requires a local SU(2)$\times$SU(2)
symmetry, which can be introduced by writing $U(x)$ as a product of three
unitary matrices\cite{bando2,meissner,bky},
$$U(x)=\xi_L(x)^\dagger \xi_M(x) \xi_R(x). \eqno(4.2)$$
In this case, the new variables are symmetric under
$$\xi_R(x)\to h_R(x)\xi_R(x)g_R^\dagger$$
$$\xi_L(x)\to h_L(x)\xi_L(x)g_L^\dagger$$
$$\xi_M(x)\to h_L(x)\xi_M(x)h_R^\dagger(x),\eqno(4.3)$$
where $h_{L,R}(x)$ are SU(2) matrices with arbitrary $x$-dependence. The
freedom to make space-time dependent rotations of $\xi_{r,l,m}(x)$ in this way
provides the local SU(2)$\times$SU(2) symmetry of this scheme. This can be
reduced to the simpler case of a local SU(2) symmetry by setting $\xi_M(x)=1$.

One can always to choose to work in the unitary gauge where
$$\xi_R(x)=\xi_L^\dagger(x)=u(x),\qquad \xi_M(x)=1, \eqno(4.4)$$
for all $x$. The symmetry (4.3) then reduces to the usual nonlinear realisation
of chiral symmetry\cite{wein,ccwz}, as in (2.23), where the $x$ dependence
of $h(x)=h_R(x)=h_L(x)$ is no longer arbitrary but is given in terms of the
pion fields. This gauge fixing thus provides the basis for translating between
the hidden-gauge and WCCWZ formalisms.

In this approach, spin-1 fields are introduced as gauge bosons of this
artificial local symmetry. Right- and left-handed gauge fields transform under
the symmetry as, respectively,
$$\widehat X_\mu(x)\to h_R(x)\widehat X_\mu(x)h_R^\dagger(x)+{i\over 
{\sqrt 2}g}h_R(x)\partial_\mu h_R^\dagger(x)$$
$$\widehat Y_\mu(x)\to h_L(x)\widehat Y_\mu(x)h_L^\dagger(x)+{i\over 
{\sqrt 2}g}h_L(x)\partial_\mu h_L^\dagger(x),\eqno(4.5)$$
where I use hats to distinguish the hidden-gauge spin-1 fields from those of
the WCCWZ approach. The corresponding gauge-covariant field strengths are 
$$\widehat X_{\mu\nu}= \partial_\mu \widehat X_\nu-\partial_\nu \widehat X_\mu
-i{\sqrt 2}g[\widehat X_\mu,\widehat X_\nu]$$
$$\widehat Y_{\mu\nu}= \partial_\mu \widehat Y_\nu-\partial_\nu \widehat Y_\mu
-i{\sqrt 2}g[\widehat Y_\mu,\widehat Y_\nu].\eqno(4.6)$$
It is usually more convenient to work in terms of the vector and axial fields,
$\widehat V_\mu=(\widehat X_\mu+\widehat Y_\mu)/\sqrt 2$ and $\widehat
A_\mu=(\widehat X_\mu-\widehat Y_\mu)/\sqrt 2$. The field strengths for these
are 
$$\widehat V_{\mu\nu}= \partial_\mu \widehat V_\nu-\partial_\nu \widehat V_\mu
-ig[\widehat V_\mu,\widehat V_\nu]-ig[\widehat A_\mu,\widehat A_\nu]$$
$$\widehat A_{\mu\nu}= \partial_\mu \widehat A_\nu-\partial_\nu \widehat A_\mu
-ig[\widehat V_\mu,\widehat A_\nu]-ig[\widehat A_\mu,\widehat V_\nu].
\eqno(4.7)$$

The gauge-covariant first derivatives of the pion fields are 
$$R_\mu= -i\left[(\partial_\mu\xi_L)\xi^\dagger_L
-i{\sqrt 2}g\widehat X_\mu\right]$$
$$L_\mu= -i\left[(\partial_\mu\xi_R)\xi^\dagger_R-i{\sqrt 2}g\widehat 
Y_\mu\right]$$
$$M_\mu= -i\left[(\partial_\mu\xi_M)\xi^\dagger_M+i{\sqrt 2}g\xi_M\widehat 
X_\mu\xi^\dagger_M-i{\sqrt 2}g\widehat Y_\mu\right].\eqno(4.8)$$
Of these, $R_\mu$ transforms covariantly under the right-handed local symmetry,
$L_\mu$ and $M_\mu$ under the left-handed. A general gauge-invariant Lagrangian
in this approach consists of all terms that can be constructed out of traces
of products of $R_\mu$, $L_\mu$, $M_\mu$, $\widehat X_{\mu\nu}$, $\widehat
Y_{\mu\nu}$, and their covariant derivatives, and that are symmetric under
parity. Factors of $\xi_M$ and $\xi^\dagger_M$ should be inserted between
right- and left-covariant quantities. Writing out explicitly only terms of
second order in the pion field gradients (which also provide mass terms for the 
heavy mesons) and the vector-meson kinetic terms, one has the
Lagrangian\cite{bando2,meissner,bky}
$${\cal L}= {af_\pi^2\over 4}\langle (L_\mu+\xi_M R_\mu \xi^\dagger_M)^2\rangle 
+{bf_\pi^2\over 4}\langle (L_\mu-\xi_M R_\mu \xi^\dagger_M)^2\rangle$$
$$+{cf_\pi^2\over 4}\langle M_\mu M^\mu\rangle +{df_\pi^2\over 4}\langle 
(L_\mu-\xi_M R_\mu \xi^\dagger_M-M_\mu)^2\rangle$$
$$-{1\over 2}\langle \widehat X_{\mu\nu}\widehat X^{\mu\nu}+\widehat Y_{\mu\nu}
\widehat Y^{\mu\nu}\rangle +\cdots.\eqno(4.9)$$

In the unitary gauge defined by (4.4), the transformation
properties of the spin-1 fields are
$$\widehat V_\mu \to h \widehat V_\mu h^\dagger+{i\over g}h\partial_\mu 
h^\dagger$$
$$\widehat A_\mu \to h \widehat A_\mu h^\dagger,\eqno(4.10)$$
where $h$ is the compensating SU(2) rotation of the nonlinear realisation
of chiral symmetry, (2.23). The covariant gradients of (4.8) reduce to
$$R_\mu=iu^\dagger\partial_\mu u-g(\widehat V_\mu+\widehat A_\mu)$$
$$L_\mu=iu\partial_\mu u^\dagger-g(\widehat V_\mu-\widehat A_\mu)$$
$$M_\mu=2g\widehat A_\mu.\eqno(4.11)$$
Since these can be combined to give
$$R_\mu+L_\mu=2(i\Gamma_\mu-g\widehat V_\mu)$$
$$R_\mu-L_\mu=u_\mu-2g\widehat A_\mu,\eqno(4.12)$$
we find that $\widehat V_\mu$ always appears in the combination 
$$V_\mu=\widehat V_\mu-{i\over g}\Gamma_\mu. \eqno(4.13)$$
This transforms homogeneously under the nonlinear chiral rotation, as can be
seen from (2.27) and (4.10).

In the unitary gauge we can therefore change variables to the vector field
$V_\mu$ of (4.13) to obtain a Lagrangian of the WCCWZ type. (The axial
field already transforms homogeneously in this gauge, (4.10).) With the aid
of (2.31), the field strengths (4.7) can be expressed in terms of
the new fields as
$$\widehat V_{\mu\nu}= V_{\mu\nu}+{i\over 4g}[u_\mu,u_\nu]-ig[V_\mu,V_\nu]
-ig[A_\mu,A_\nu]$$
$$\widehat A_{\mu\nu}= A_{\mu\nu}-ig[V_\mu,A_\nu]-ig[A_\mu,V_\nu],\eqno(4.14)$$
where the covariant field gradients are defined in (3.1) above. Terms involving
higher gauge-covariant derivatives can be rewritten in terms of the covariant
derivative (2.29) using
$$\widehat D_\mu=\partial_\mu-ig[\widehat V_\mu,\ ]=\nabla_\mu-ig[V_\mu,\ ].
\eqno(4.15)$$
Each term of the general hidden-gauge Lagrangian in the unitary gauge has a
corresponding term in the general WCCWZ Lagrangian, where $\widehat V_\mu
-i\Gamma_\mu/g$ has been replaced by $V_\mu$, $\widehat D_\mu$ by $\nabla_\mu$,
$\widehat V_{\mu\nu}$ by $V_{\mu\nu}$, and $\widehat A_{\mu\nu}$ by
$A_{\mu\nu}$. The coupling constants will not be identical but, if one takes
account of Eqs.~(4.14,15), there is a well defined way to convert from one
approach to the other. This generalises Georgi's observation\cite{georgi} of
the equivalence of the two formalisms to the case of axial as well as vector
fields. 

An important feature to note is that the gauge coupling constant $g$ of the
hidden-gauge approach does not appear in the WCCWZ approach. Indeed
hidden-gauge Lagrangians with different values of $g$, and containing different
higher-order interactions, can be equivalent to the same WCCWZ theory. This
should not be too surprising: the local symmetry is not physical but arises
from a particular choice of field variables in (4.1,2) and hence the
corresponding coupling is not a physical quantity. The significance of $g$
becomes clearer if one starts from a WCCWZ Lagrangian and converts it into a
hidden-gauge one using (4.13) in reverse. Any value of $g$ can be used in
(4.13) to define a new vector field $\widehat V_\mu$ and the resulting
Lagrangian will have the form of a hidden-gauge theory in the unitary gauge.
Different choices of $g$ therefore correspond to different choices of vector
field. The value of $g$ should thus be fixed by considerations of
calculational convenience, for example the elimination of certain types of
term from the effective Lagrangian.

To explore this equivalence in more detail, let us examine a specific 
hidden-gauge theory. The example considered is the most commonly used
hidden-gauge model, introduced by Bando {\it et al.}\cite{bando1}. This
contains a vector but no axial field and so is invariant under the diagonal
SU(2) subgroup of the local symmetry only. Its Lagrangian has the form
$${\cal L}={f_\pi^2\over 4}\langle (L_\mu-R_\mu)^2\rangle 
+{af_\pi^2\over 4}\langle (L_\mu+R_\mu)^2\rangle 
-{1\over 2}\langle \widehat V_{\mu\nu}\widehat V^{\mu\nu}\rangle ,\eqno(4.16)$$
where 
$$R_\mu= -i\left[(\partial_\mu\xi_L)\xi^\dagger_L
-ig\widehat V_\mu\right]$$
$$L_\mu= -i\left[(\partial_\mu\xi_R)\xi^\dagger_R-ig\widehat V_\mu\right].
\eqno(4.17)$$
In the unitary gauge this becomes
$${\cal L}={f_\pi^2\over 4}\langle u_\mu u^\mu\rangle 
+af_\pi^2\langle (i\Gamma_\mu-g\widehat V_\mu)^2\rangle 
-{1\over 2}\langle \widehat V_{\mu\nu}\widehat V^{\mu\nu}\rangle, \eqno(4.18)$$
showing that the $\rho$ mass is given in terms of the parameter $a$ by
$$m_V^2=ag^2f_\pi^2. \eqno(4.19)$$

The replacement of the covariant derivatives (3.1) by gauge-covariant ones
(4.7), means that the Weinberg-Tomozawa piece of $\pi\rho$ scattering no
longer appears in the kinetic term for the $\rho$ in the Lagrangian. Instead
it is generated entirely from $\rho$ exchange. The fact that this leads to the
current-current interaction (2.13) can be seen from the $\rho\pi\pi$ coupling
contained in (4.18),
$$-2igaf_\pi^2\langle \widehat V^\mu \Gamma_\mu\rangle ={1\over 2}ag
\widehat{\hbox{\bf V}}^\mu\cdot\hbox{\boldmath$\pi$}\wedge\partial_\mu
\hbox{\boldmath$\pi$}+{\cal O}(\pi^4), \eqno(4.20)$$
and the $3\rho$ coupling
$$2ig\langle (\partial_\mu\widehat V_\nu-\partial_\nu\widehat V_\mu)
[\widehat V^\mu,\widehat V^\nu]\rangle =-g\widehat{\hbox{\bf V}}^\mu\cdot
\widehat{\hbox{\bf V}}^\nu\wedge\partial_\mu \widehat{\hbox{\bf V}}_\nu.
\eqno(4.21)$$
Moreover the factor $a$ in the $\rho\pi\pi$ coupling cancels with that from
the $\rho$ mass (4.19) in the denominator. The low-energy theorem is thus
satisfied independently of the value of the $\rho$ mass.

Using (4.13,14) the Lagrangian (4.18) can be expressed in the equivalent WCCWZ 
form (3.4), with the following values for the coupling constants:
$$g_1={1\over 2g},\quad g_2=2g,\quad g_3={1\over 4g^2},\quad g_4=1,\quad 
g_5=4g^2. \eqno(4.22)$$
The couplings in this model thus satisfy the relations 
$$g_3=g_1^2,\quad g_4=g_1g_2,\quad g_5=g_2^2, \eqno(4.23)$$
which arise from assuming resonance saturation by the $\rho$-meson, as
discussed in Sec.~III. The other condition that defines the model is a
relation between the $\rho\pi\pi$ and $3\rho$ couplings,
$$g_1={1\over g_2}. \eqno(4.24)$$
These relations (4.23,24) allow the three- and four-point couplings to be
combined into a kinetic term for the vector field with a Yang-Mills form.
Note that all of these relations hold for any value of the $\rho$ mass (or
equivalently of the parameter $a$).

The replacement of derivatives by gauge-covariant ones involving the external
fields generates all the usual minimal electromagnetic couplings. The presence
in the connection (2.38) of a term linear in the vector field $v_\mu$ means
that the mass term in (4.18) develops a $\gamma\rho$ mixing piece. The
presence of such a momentum-independent mixing term makes the hidden-gauge
formalism an especially convenient one for embodying the idea of vector
dominance. The coefficient of this mixing term is $agf_\pi^2$. By comparing
this with the $\rho\pi\pi$ coupling (4.20), we see that the KSFR relation
(2.16) is satisfied\cite{bando2,bando3}, as it should be since the Lagrangian
has been constructed to respect chiral symmetry. It is unsurprising that this
low-energy theorem continues to hold when loop corrections are
included\cite{hy,hky}.

Other couplings can obtained from the corresponding WCCWZ forms by making the
replacements (4.13,14) above. For example, an additional, momentum-dependent
$\gamma\rho$ mixing can be described by a term of the form $\langle\widehat
V_{\mu\nu}F_+^{\mu\nu}\rangle$.

As an example, consider an extension of the model (4.16) with only the minimal
replacement of derivatives. In converting this into its WCCWZ equivalent, we
need to note that the presence of the field tensor in (2.40) produces an
additional term in the relation corresponding to (4.14),
$$\widehat V_{\mu\nu}= V_{\mu\nu}+{i\over 4g}[u_\mu,u_\nu]
+{i\over g}F_{+\mu\nu}-ig[V_\mu,V_\nu]. \eqno(4.25)$$
This leads to a $\gamma\rho$ mixing of the form (3.7) together with the terms
given in (3.10) and an anomalous magnetic moment for the $\rho$. The couplings
constants for these are all related to $g$, 
$$f_1={1\over g},\quad f_2={1\over 2g^2},\quad f_3={1\over g^2}. \eqno(4.26)$$
The model thus satisfies the resonance saturation conditions (3.13,14) for
the electromagnetic couplings, in addition to those of (4.23).

There is one other relation between the coupling constants that defines this
model. This is a connection between the $\rho\pi\pi$ and $\gamma\rho$
couplings\cite{ecker2},
$$f_1=2g_1, \eqno(4.27)$$
which does not follow from resonance saturation. The factor of two in this
relation makes it very reminiscent of the KSFR relation (2.16). Although it
refers to the momentum-dependent couplings of the WCCWZ formalism and not
those appearing in (2.16), this relation is nonetheless very closely related
to the KSFR one, as we shall see below.

Vector dominance of the pion electromagnetic form factor is obtained if the
photon coupling to the pion current in the first term of (4.18) cancels with
the corresponding piece of the second term. The photon then couples to the
pion only through a virtual $\rho$. This will happen if the $\rho$ mass
satisfies
$$m_V^2=2g^2f_\pi^2, \eqno(4.28)$$
which is just the KSFR relation in its second, vector-dominance form (2.18).
In terms of the parameters of Bando {\it et al.}\cite{bando1,bky} this
corresponds to $a=2$. Vector dominance in the couplings of the photon to all
hadrons requires universal coupling of the $\rho$ meson to the conserved
isospin current. From the expressions for the $\rho\pi\pi$ and $3\rho$
couplings (4.20,21), we can see that for $a=2$ these have the same strength
and so this model also embodies universal coupling of the $\rho$ to itself and
to the pion.

In looking at predictions of the model (4.16), consequences of the hidden-gauge
choice of field should not be confused with those arising from the relations
between the coupling constants (4.23,24,26,27). The former controls merely the
form of off-shell extrapolations of those amplitudes. The latter lead to
relations between amplitudes for physical processes, and are of course
specific to the choice of Lagrangian. For example the relation (4.24) between
the $\rho\pi\pi$ and $3\rho$ couplings can be removed without violating the
hidden-gauge invariance by adding a term of the form $\langle \widehat
V_{\mu\nu} \left[R^\mu+L^\mu,R^\nu+L^\nu\right]\rangle$ to the Lagrangian
(4.16). This is invariant under the same local SU(2) symmetry as the rest of
the Lagrangian. It provides an additional contribution to the $3\rho$ coupling
beyond that in the kinetic term. After gauge-fixing and change of variables,
such a term would lead to an equivalent WCCWZ Lagrangian that would not
satisfy (4.24).

To see why the specific choice of field in (4.18) is particularly convenient,
remember that the $\rho\pi\pi$ of the corresponding WCCWZ Lagrangian (3.5) is
of order $p^3$. Also the $\gamma\rho$ mixing (3.8) is of order $p^2$. In
contrast, when the model is expressed in hidden-gauge form, the leading
$\rho\pi\pi$ coupling (4.20) is of first order in the momenta of the particles
involved and the $\gamma\rho$ mixing is independent of momentum. Using the
field defined by (4.13) with the constant $g$ given by
$$g={1\over 2g_1} \eqno(4.29)$$ 
eliminates any ${\cal O}(p^3)$ $\rho\pi\pi$ term from the hidden-gauge 
Lagrangian. This happens even if the initial WCCWZ Lagrangian contains other,
higher-derivative $\rho\pi\pi$ couplings. The advantage of the hidden-gauge
choice of field with this $g$ is that any corrections to the leading
$\rho\pi\pi$ of (4.20) are of order $p^5$ or higher. Provided
$m_V$ is small compared with the scale at which physics beyond the $\pi\rho$
Lagrangian becomes significant, the momentum dependence of the effective
$\rho\pi\pi$ coupling should be small in the hidden-gauge representation. 

Moreover, if the $\gamma\rho$ mixing strength of the WCCWZ Lagrangian satisfies
the condition (4.27) then the same choice of field also eliminates any ${\cal 
O}(p^2)$ mixing term from the hidden-gauge Lagrangian. Corrections to it are
thus at least of order $p^4$. Hence the leading corrections to both of the
coupling constants that appear in (2.16) are of order $p^4$ instead of $p^2$
in this model. The model (4.16) thus embodies the empirical observation that
the KSFR relation in its first version (2.16), which relates the $\rho\pi\pi$
coupling and $\gamma\rho$ mixing at zero four-momentum, is actually well
satisfied by the values for on-shell $\rho$ mesons.\footnote{I am assuming 
here that the order $p^4$ corrections are small compared to the leading terms
in the expansion. In principle the KSFR relation could also be satisfied if the
higher-order contributions are large but remain in the same 2:1 ratio as those
in (2.16) and (4.27).}

The relation (4.27) between the couplings of the equivalent WCCWZ theory is
what makes it possible for the KSFR relation to be satisfied on-shell as well
as at zero four-momentum. However one should remember that the manipulations
here shed no light on the origin of this. Like vector dominance, it remains
an phenomenologically successful assumption that arises neither from chiral
symmetry nor from the assumption of resonance saturation.

\section{Massive Yang-Mills}

The massive Yang-Mills approach\cite{gg,syracuse,meissner} is based on vector
and axial fields that transform linearly under the SU(2)$\times$SU(2)
symmetry, as in (2.22). The Lagrangian for these is chosen to contain kinetic
terms of the Yang-Mills form, including three- and four-point interactions.
The couplings of the spin-1 fields to pions are also chosen to have a
gauge-invariant form, ensuring universal coupling of the $\rho$ and allowing
photons to be coupled in a way consistent with vector dominance. Although the
interaction terms respect a local SU(2)$\times$SU(2) symmetry, the full theory
does not since it also includes mass terms for the spin-1 mesons.

A simple massive Yang-Mills Lagrangian, which illustrates the features of the
approach, consists of the gauged sigma model (a nonlinear version of the model
used by Gasiorowicz and Geffen\cite{gg})
$${\cal L}={f_0^2\over 4}\langle\widetilde D_\mu U(\widetilde D_\mu U)^\dagger
\rangle-{1\over 2}\langle\widetilde X_{\mu\nu}\widetilde X^{\mu\nu}
+\widetilde Y_{\mu\nu}\widetilde Y^{\mu\nu}\rangle$$
$$+m_V^2\langle \widetilde X_\mu\widetilde X^\mu
+\widetilde Y_\mu\widetilde Y^\mu\rangle, \eqno(5.1)$$
where
$$\widetilde D_\mu U=\partial_\mu U+i\sqrt 2\widetilde gU\widetilde X_\mu
-i\sqrt 2\widetilde g\widetilde Y_\mu U, \eqno(5.2)$$
and the field strengths $\widetilde X_{\mu\nu}$, $\widetilde Y_{\mu\nu}$
are defined analogously to those in (4.6).

In this approach, low-energy theorems cannot simply be read off from terms in
the Lagrangian; they are obtained from combinations of a number of pieces. A
further complication is the presence of a $\pi a_1$ mixing term which can be
removed by an appropriate shift in the definition of the axial
field\cite{gg,ecker2}. The gauge-like couplings of the $\rho$ mean that
($t$-channel) $\rho$ exchange contributes to the Weinberg-Tomozawa term for
$\pi\rho$ scattering. However one has also to include pieces with intermediate
($s$- and $u$-channel) $\pi$ and $a_1$ states to satisfy the low-energy
theorem. The isospin-symmetric pion scattering amplitude, which should vanish
for at threshold in the chiral limit ({\it cf.}\cite{wein}), is also rather
complicated. It contains a momentum-independent contribution from the pion
kinetic term of (5.1), but this is exactly cancelled by contributions
involving intermediate $\pi$ and $a_1$ states\cite{plant}. All of this shows
that, in the massive Yang-Mills formalism, one has to take care to include all
possible contributions to any process as otherwise low-energy theorems will be
violated.

The massive Yang-Mills theory can be converted into an equivalent WCCWZ one by
using $u(x)$ to construct spin-1 fields that transform under the nonlinear
realisation of chiral symmetry as described in Sec.~IIC. The new fields are 
given by (2.33) and the kinetic terms can be expressed in terms of their 
covariant derivatives, defined as in (3.1), using
$$\widetilde X_{\mu\nu}=u^\dagger\Biggl[X_{\mu\nu}+{i\over 2}[u_\mu,X_\nu]
-{i\over 2}[u_\nu,X_\mu]-i\sqrt 2\widetilde g[X_\mu,X_\nu]\Biggr]u$$
$$\widetilde Y_{\mu\nu}=u^\dagger\Biggl[Y_{\mu\nu}-{i\over 2}[u_\mu,Y_\nu]
+{i\over 2}[u_\nu,Y_\mu]-i\sqrt 2\widetilde g[Y_\mu,Y_\nu]\Biggr]u.\eqno(5.3)$$
In terms of $u_\mu$ and these fields, the pion kinetic term can be written
$$\langle\widetilde D_\mu U(\widetilde D_\mu U)^\dagger\rangle=\langle[u_\mu
-\sqrt 2\widetilde g(X_\mu-Y_\mu)]^2\rangle.\eqno(5.4)$$
This contains the $\pi a_1$ mixing term mentioned above. To remove this, it
is convenient to define WCCWZ vector and axial fields by
$$V_\mu={1\over \sqrt 2}(X_\mu+Y_\mu)$$
$$A_\mu={1\over \sqrt 2}(X_\mu-Y_\mu)-{\widetilde g f_0^2\over 2m_A^2}u_\mu.
\eqno(5.5)$$

The kinetic terms for the spin-1 fields can then be expressed in terms of
$V_\mu$ and $A_\mu$ and their covariant derivatives (3.1), making use of 
(2.30). The Lagrangian (5.1) then takes the form
$${\cal L}={f_\pi^2\over 4}\langle u_\mu u^\mu\rangle+m_V^2\langle V_\mu 
V^\mu\rangle+m_A^2\langle A_\mu A^\mu\rangle$$
$$-{1\over 2}\Bigg\langle\Bigg\{V_{\mu\nu}-i\widetilde g[V_\mu,V_\nu]
-i\widetilde g[A_\mu,A_\nu]$$
$$+i{\textstyle{1\over 2}}Z^2\Bigl([u_\mu,A_\nu]-[u_\nu,A_\mu]\Bigr)$$
$$+{i\over 4\widetilde g}(1-Z^4)[u_\mu,u_\nu]\Bigg\}^2\Bigg\rangle$$
$$-{1\over 2}\Big\langle\Big\{A_{\mu\nu}-i\widetilde g[V_\mu,A_\nu]
-i\widetilde g[A_\mu,V_\nu]$$
$$+i{\textstyle{1\over 2}}Z^2\Bigl([u_\mu,V_\nu]
-[u_\nu,V_\mu]\Bigr)\Big\}^2\Big\rangle. \eqno(5.6)$$
where 
$$Z^2=1-{\widetilde g^2 f_0^2\over m_A^2}=1-{\widetilde g^2 f_\pi^2\over 
m_V^2}, \eqno(5.7)$$
and the physical pion decay constant is given by\cite{meissner}
$$f_\pi^2=f_0^2Z^2, \eqno(5.8)$$
and the $a_1$ mass by
$$m_A^2=m_V^2/Z^2. \eqno(5.9)$$

Although I have demonstrated the equivalence here for only the theory defined
by (5.1), it is general. Any massive Yang-Mills Lagrangian can be expressed in
WCCWZ form using (5.2--5). Conversely, any WCCWZ Lagrangian can be converted
into an equivalent massive Yang-Mills theory by inverting these changes of
variable. Of course the resulting Lagrangian can contain many terms beyond
those present in (5.1), including many non-gauge-invariant interactions.
Combined with the results of Section III, this reproduces and generalises the
well-known equivalence of the hidden-gauge and massive Yang-Mills
formalisms\cite{schechter,mz,yamaw,meissner}.

By comparing the terms in the Lagrangian (5.6) with the corresponding ones in
(3.4), we can see that the couplings satisfy the relations (4.23) arising from
assuming resonance saturation of the four-point interactions. This is similar
to the hidden-gauge theory defined by (4.16). The two theories are thus closely
related although obviously not identical: the massive Yang-Mills one contains
an axial as well as a vector field, and its $\rho\pi\pi$ and $3\rho$ couplings
do not satisfy (4.24). The latter is a consequence of the the additional
momentum-dependent $\rho\pi\pi$ couplings that appear after diagonalising in
the $\pi a_1$ sector. One can always cancel out this momentum dependence by
adding extra nonminimal terms to the massive Yang-Mills
Lagrangian\cite{syracuse}.\footnote{One can also force the theory into exact
equivalence with the one of Bando {et al.}\cite{bando1} by imposing a suitable
constraint on the axial field\cite{ks,meissner}. This can be seen using the
WCCWZ form (5.6): if one demands that $A_\mu=(Z^2/2g)u_\mu$ then one is left
with the WCCWZ equivalent of (4.16).} The resulting massive Yang-Mills
Lagrangian is then exactly equivalent to the hidden-gauge one involving
axial as well as vector mesons introduced in Ref.~\cite{bando3}.

Electromagnetic couplings can be included in the usual way by replacing
derivatives with gauge-covariant derivatives involving the external fields.
Nonminimal terms can be constructed, with a little care to ensure that they
respect chiral invariance. For example, $\gamma\rho$ mixing can be obtained
from a combination of two kinetic mixing terms, $\langle X_{\mu\nu}(v^{\mu\nu}
+a^{\mu\nu})+Y_{\mu\nu}(v^{\mu\nu}-a^{\mu\nu})\rangle$ and
$\langle X_{\mu\nu}U^\dagger(v^{\mu\nu}-a^{\mu\nu})U
+Y_{\mu\nu}U(v^{\mu\nu}+a^{\mu\nu})U^\dagger\rangle$.

Vector dominance can be realised if the $\rho$ mass satisfies the second form
of the KSFR relation (2.18). In this case $Z=1/2$ and the $\rho$ and $a_1$ 
masses satisfy the Weinberg relation $m_A=\sqrt 2m_V$\cite{wrel}. Also, with
this value of the $\rho$ mass, $\rho$ exchange generates the whole of the
Weinberg-Tomozawa term.

For completeness, I should mention the approach suggested by Brihaye,
Pak and Rossi\cite{bpr} and investigated further by Kuraev, Silagadze and
coworkers\cite{kuraev,esk}. This is based on a Yang-Mills-type coupling of the
$\rho$, as in the Lagrangian (5.1), but without a chiral partner $a_1$-field.
Simply omitting the axial field from that Lagrangian leaves a theory that is
not chirally symmetric. However as described in Ref.\cite{bpr} additional
counterterms can be added to that Lagrangian to ensure that low-energy theorems
arising from chiral symmetry are maintained.

Such a theory can be generated by taking a hidden-gauge Lagrangian, such
as that of (4.18), and reversing the procedure above for converting fields
that transform linearly under chiral symmetry into ones that transform 
nonlinearly. Specifically one can define a new vector field $\widetilde V_\mu$,
related to the hidden-gauge field $\widehat V_\mu$ by
$$\widehat V_\mu={1\over 2}\left(u^\dagger \widetilde V_\mu u
+u\widetilde V_\mu u^\dagger\right). \eqno(5.10)$$
Note that this $\widetilde V_\mu $ has no axial partner and so does not
transform in any simple way under chiral transformations. By adding and
subtracting suitable terms, similar to those in (5.4), the pion kinetic
term can be converted to a form involving gauge-covariant derivatives. The full
Lagrangian is then
$${\cal L}={f_\pi^2\over 4}\langle \widetilde D_\mu U(\widetilde D_\mu 
U)^\dagger\rangle+{\widetilde gf_\pi^2\over 2}\langle\widetilde V_\mu
(uu_\mu u^\dagger-u^\dagger u_\mu u)\rangle-{\widetilde g^2f_\pi^2\over 4}
\langle (u^\dagger V_\mu u-u V_\mu u^\dagger)^2\rangle$$
$$+af_\pi^2\langle (i\Gamma_\mu-g\widehat V_\mu)^2\rangle
-{1\over 2}\langle \widehat V_{\mu\nu}\widehat V^{\mu\nu}\rangle. 
\eqno(5.11)$$
By choosing the coupling $\widetilde g$ to be related to the parameters
of the hidden-gauge Lagrangian by
$$\widetilde g={ag\over 2}, \eqno(5.12)$$
one can ensure that the ${\cal O}(p)$ $\rho\pi\pi$ coupling (4.20) in the
fourth term of (5.11) is cancelled by that in the second term. The 
$\rho\pi\pi$ coupling is then given entirely by the gauge-covariant
derivatives in the first term of (5.11). 

The resulting theory has a Yang-Mills structure for the $\rho$ kinetic energy
and $\rho\pi\pi$ couplings, together with a $\rho$ mass term and a number of
additional couplings. These extra couplings are required if the low-energy
theorems of chiral symmetry, such as (2.11), are to be satisfied. They include
the counterterms discussed in Refs.\cite{bpr,kuraev} together with many
others. For example the third term of (5.11) contains a momentum-independent
$\rho\rho\pi\pi$ coupling. This term is omitted in the calculations of Kuraev
{\it et al.}\cite{kuraev,esk} but it is needed to cancel out a corresponding
piece of the pion kinetic term, which would otherwise give a nonzero amplitude
for $\pi\rho$ scattering at threshold in the chiral limit.

The consequences of failing to include all the necessary counterterms in this
approach are illustrated by the calculations of the rare decays
$\rho\rightarrow 2\pi^+\pi^-$ and $\rho\rightarrow 2\pi^0\pi^+\pi^-$, which
have been proposed as possible tests of effective Lagrangians for vector
mesons \cite{bgp,esk}.\footnote{It should be clear that the formal equivalence 
between the various schemes described here means that there is no way to 
discriminate between them empirically. Nonetheless one might still hope to test
assumptions about the values of some of the other couplings in particular
Lagrangians, for example the $3\rho$ vertex\cite{pb}.} The decay rates obtained
from various Lagrangians that respect the constraints of chiral
symmetry\cite{pb} are at least an order of magnitude smaller than those from
Lagrangians that violate some of these constraints\cite{bgp,esk}.

\section{Tensor fields}

The other main formalism for inclusion of spin-1 mesons differs from those we
have seen so far in that it uses antisymmetric tensor fields to describe these
particles\cite{ecker1} (see\cite{pp} for the extension to the anomalous
sector). The field describing the $\rho$ and $a_1$ mesons are denoted here by
$\overline V_{\mu\nu}$ and $\overline A_{\mu\nu}$, and these transform
homogeneously under the nonlinear realisation of chiral symmetry in Sec.~IIC.
The mixed space-time components of these fields, $\overline V_{0i}$ and 
$\overline A_{0i}$, describe the physical degrees of freedom. The others, like
the time components of the vector fields, should be eliminated by constraints.
The simplest Lagrangian of this form for pions and $\rho$ mesons coupled to
external fields is\cite{ecker1,ecker2}
$${\cal L}={f_\pi^2\over 4}\langle u_\mu u^\mu\rangle -{1\over 2}\langle 
\nabla^\lambda\overline V_{\lambda\mu}\nabla_\nu\overline V^{\nu\lambda}\rangle 
+{m_V^2\over 2}\langle\overline V_{\mu\nu}\overline V^{\mu\nu}\rangle 
+{i\over 2} G_1 \langle \overline V_{\mu\nu} [u^\mu,u^\nu]\rangle  
+F_1\langle\overline V_{\mu\nu}F_+^{\mu\mu}\rangle. \eqno(6.1)$$

The covariant derivatives $\nabla_\mu$ in the $\rho$ kinetic term mean that 
the Weinberg-Tomozawa amplitude can be obtained directly from that term, as in
the WCCWZ formalism. The coupling terms in (6.1) are analogous to terms in 
(3.4,8) but differ in that they involve one less power of momentum. Their
contributions to decays of on-shell $\rho$ mesons are identical if the coupling
constants are related by\cite{ecker2}
$$G_1=m_Vg_1,\quad F_1=m_Vf_1. \eqno(6.2)$$
Moreover the smaller momentum dependence of these interactions means that
amplitudes calculated in this scheme satisfy the constraints arising from
unitarity and QCD combined with resonance saturation. There is thus no need
to supplement the Lagrangian (6.1) with four-point couplings of the type
discussed in Sec.~III. Hence, if resonance saturation is assumed, this scheme
provides a particularly economical expression for the Lagrangian.

A complete proof of the equivalence of the WCCWZ and tensor-field schemes does
not exist at present. For any term in the general WCCWZ Lagrangian one can
write down an analogous one for the tensor fields. This is obvious for any
interaction terms constructed out of the antisymmetric covariant derivatives
(3.1), which can simply be replaced by the tensor fields. Other interactions
can be generated by making the replacements
$$V_\mu\rightarrow-{1\over m_V}\nabla^\lambda\overline V_{\lambda\mu},\quad
A_\mu\rightarrow-{1\over m_V}\nabla^\lambda\overline A_{\lambda\mu}. 
\eqno(6.3)$$
This replacement should also be used in any terms involving the symmetric 
covariant derivatives\cite{bm}. However beyond order $p^4$ any demonstration 
of the exact equivalence of the resulting Lagrangian to the original is
complicated by the need to enforce constraints on the fields so that they
describe only physical degrees of freedom. As result the equivalence of the
two schemes has been shown only for terms up to order $p^4$ in the WCCWZ 
Lagrangian\cite{ecker2,bm,bp}.

The substitution just described is purely a book-keeping device for
enumerating the terms in the general tensor-field Lagrangian. Unlike the
changes of variables described for the vector fields, it does not provide a
means to convert a particular WCCWZ Lagrangian directly into tensor-field form
or {\it vice versa}. In practice what is usually done is to integrate out the
spin-1 fields from both approaches and then to compare the local terms in the
resulting effective actions for pions only\cite{ecker2,bm}. As an illustration
of this, consider the Skyrme term. Integrating out the $\rho$ in the WCCWZ case
generates a contribution to the coefficient of this term of second-order in
$g_1$, in addition to the $g_3$ originally in the Lagrangian (5.4). The net
coefficient should match with that in the tensor-field case, where it arises
purely from any Skyrme term in the Lagrangian. If resonance saturation holds
then $g_3=g_1^2$ and the two contributions exactly cancel in the WCCWZ case.
There is then no explicit Skyrme term in the equivalent tensor-field Lagrangian
(6.1).

Recently, a path integral approach to re-expressing a vector-field theory in
terms of tensor fields has been described in\cite{bp}. This is inspired by the
dual transformation of gauge theories\cite{dual} and its starting point is a
change of variables corresponding to (6.3) in the path integral. This procedure
does generate, for example, the change in the coefficient of the Skyrme term
and in principle it could be used to translate a Lagrangian from WCCWZ to
tensor-field formalism. However beyond the lowest-order terms of the Lagrangian
(3.4,8) there are practical problems that arise from the constraints on the 
fields.

\section{Symmetry breaking}

So far I have concentrated on the chirally symmetric parts of the Lagrangians
in the various approaches. These can be straightforwardly extended to 
incorporate the effects of explicit symmetry breaking by the current quark
masses. Such Lagrangians with vector mesons have been the subject of much
recent discussion in the context of charge-symmetry breaking in the $NN$
interaction. These effects arise because the difference between the up- and
down-quark current masses breaks isospin as well as chiral symmetry. Much of
this interest has focussed on the momentum-dependence of the mixing between the
$\rho$ and $\omega$ mesons (see\cite{oconn,malt} and references therein). The
forms of possible $\rho\omega$ mixing vertices are analogous to those for
$\gamma\rho$ mixing, with one exception: the absence of gauge invariance means
that both mass and kinetic mixing terms are permitted in any formalism.

For example, such effects can be included in a WCCWZ Lagrangian with terms of
the form
$${\cal L}={f_\pi^2\over 4}\langle u_\mu u^\mu\rangle -{1\over 2}\langle 
V_{\mu\nu}V^{\mu\nu}\rangle +m_V^2\langle V_\mu V^\mu\rangle 
-{1\over 4}\omega_{\mu\nu}\omega^{\mu\nu}+{1\over 2}m_\omega^2
\omega_\mu\omega^\mu$$
$$-{i\over2} g_1 \langle V_{\mu\nu} [u^\mu,u^\nu]\rangle
+{1\over 8} g_3 \langle[u_\mu,u_\nu]^2\rangle
+b_0\langle \chi_+\rangle
-2b_1\langle\chi_+ V_\mu\rangle\omega^\mu$$
$$-b_2\langle\chi_+ V_{\mu\nu}\rangle\omega^{\mu\nu}
-{i\over 2}b_3\langle\chi_+ [u_\mu,u_\nu]\rangle\omega^{\mu\nu}
+c_1\langle ([\chi_-, u_\mu]V^\mu\rangle,  \eqno(7.1)$$ 
where the explicit symmetry breaking is introduced through the $\chi_\pm$,
defined in (2.41). The first of the symmetry-breaking terms written out here
$\langle \chi_+\rangle$ produces the pion mass. The next two describe
$\rho\omega$ mixing of order $p^0$ and $p^2$. In addition there can be an
isospin-violating direct $\omega\pi\pi$ coupling of order $p^3$. Under the
assumption of resonance saturation in isospin-violating $\pi\pi$ scattering,
we find that the direct $\omega\pi\pi$ coupling is given by ({\it cf.} (3.13))
$$b_3=g_1b_2. \eqno(7.2)$$
The final term in (7.1) introduces a chiral-symmetry breaking $\rho\pi\pi$ 
coupling
of order $p$.

By making appropriate changes of variable, analogous to those used in the
context of $\gamma\rho$ mixing\cite{vmd,current,klz}, one can diagonalise
either the kinetic or mass terms of the Lagrangian to leave it in a form with
only one type of mixing. One can also go further and diagonalise the whole
free-field part of the vector-meson Lagrangian, in which case all of the
isospin violation would appear in the couplings of those mesons. (All of these
procedures would produce fields with rather complicated properties under chiral
transformations.) Alternatively the theory could be converted into
hidden-gauge form, introducing an ${\cal O} (p)$ $\omega\pi\pi$ coupling from
the change of variables in the mass-type mixing term. From all this, it should
be obvious that one can change the strengths of the various mixing terms and
isospin-violating couplings in the Lagrangian simply by changing field
variables. In calculating any symmetry-breaking amplitude, it is thus not
sufficient to know the strength of the mixing and its momentum dependence, one
must also have the corresponding symmetry-breaking couplings of the vector
mesons\cite{ghp,mow}. This point has recently been stressed by Cohen and
Miller\cite{cm}.

Within the WCCWZ formalism one could set both the momentum-dependent mixing
and the direct isospin violating couplings to zero. This would then realise
the traditional picture of charge symmetry violation arising purely from a
momentum-independent mixing, as suggested in\cite{cm}. Note that this refers
only to the tree-level mixing parameters and at this level both $\rho$ and
$\omega$ have zero width. The large width of the $\rho$ arises from its strong
coupling to two pions and several authors have pointed out that this can lead
to a significant momentum dependence of the $\rho\omega$
mixing\cite{ijl,malt2}. This suggests that the inclusion of pion loops is 
essential in calculations of charge-symmetry breaking effects.

\section{Discussion}

As I have described here, any effective theory of spin-1 mesons and pions can
be expressed in WCCWZ, hidden-gauge, massive Yang-Mills or tensor-field form.
These schemes correspond to different choices of fields for the spin-1 mesons.
The rules for transforming a theory from one form to another have been
described in detail, at least for the schemes based on vector fields. Since
they are all equivalent, the choice between them must depend on the
convenience of the corresponding Lagrangians for a specific calculation. In
discussing a particular Lagrangian, we need to be careful to distinguish
general features of the formalism used to express it from the properties of the
specific coupling constants it contains.

The massive Yang-Mills scheme is rather different from the other three since
it is based on the linear realisation of chiral symmetry. Although, correctly
treated, it satisfies chiral low-energy theorems, this is not immediately
obvious from the Lagrangian since large contributions need to be calculated
with delicate cancellations between them. This means that great care needs to
be taken if any approximations are made in this approach. In contrast, the
other schemes use the nonlinear realisation of chiral symmetry and so the
low-energy theorems are built into the forms of the terms appearing in their
Lagrangians.

The WCCWZ approach, based on vector fields that transform homogeneously under
the nonlinear realisation, imposes no preconceptions about the types of
couplings. This makes it particularly useful for comparing theories that have
been expressed in different formalisms. In addition, we have seen that the
assumptions underlying any particular theory can conveniently be expressed in
terms of relations between the coupling constants that appear in its WCCWZ
equivalent.

The hidden-gauge scheme uses fields that transform inhomogeneously under the 
nonlinear realisation and as a result involves couplings that respect a gauge
invariance. This makes a convenient formalism to use in the context of vector
dominance. 

Finally there is the tensor-field formalism. This allows for a rather compact
form for the Lagrangian if resonance saturation is assumed. Otherwise it is
rather similar to the WCCWZ approach and does not seem to have any great 
advantage over that scheme.

By rewriting various theories in WCCWZ form, we have found various
relationships amongst their coupling constants. These can be organised
hierarchically, according the underlying physical assumptions that lead to
them. First there are the inequalities like (3.6) for the four-point couplings
that follow from general principles such as existence of a vacuum or
unitarity. Under the assumption of resonance saturation by the $\rho$ and
$a_1$, these become equalities fixing the values of the four-point couplings.
Similarly resonance saturation combined with QCD expectations for the
behaviour of current-current correlators fixes the values of certain
electromagnetic couplings (3.13,14). The values for the couplings obtained
from resonance saturation agree well with those determined from ChPT. Finally
there are relations that arise from the phenomenologically successful
assumptions of vector dominance in the pion form factor (3.16) and (4.28),
universal coupling of the $\rho$ meson (4.24) and the on-shell KSFR relation
(4.27). Despite their successes, the underlying origins of these last
relations and of resonance saturation remain obscure.

An interesting feature of the hidden-gauge approach is that it involves 
a parameter $g$, the gauge coupling for the local symmetry, that has no 
counterpart in the equivalent WCCWZ Lagrangian. This coupling can be viewed
as a parameter in the hidden-gauge vector field. The conventional choice of
this field, implicitly used in all applications of the approach, is the one
that removes the ${\cal O}(p^3)$ part of the $\rho\pi\pi$ coupling. If
resonance saturation holds, this choice also removes the ${\cal O}(p^2)$
$\gamma\rho$ mixing. Hence this formalism embodies rather naturally the
observation that the on-shell $\rho$ couplings satisfy the KSFR relation
(2.16).

In summary: the various formalisms for including spin-1 mesons in effective
chiral Lagrangians are equivalent. Hence any particular theory could be 
expressed in terms of the fields of any desired scheme. The choice of scheme
is thus a question of convenience for a particular problem. If one is
interested in elucidating the physical assumptions built into some given
Lagrangian, then conversion of that theory into its WCCWZ equivalent is
recommended.

\section*{Acknowledgments}

I am grateful to D. Kalafatis for extensive discussions at the start of this
work. I would like to thank R. Plant for useful discussions and J. McGovern
for critically reading the manuscript. I am also grateful to the Institute for
Nuclear Theory, University of Washington, Seattle for its hospitality while
part of this work was carried out. This work is supported by the EPSRC and
PPARC.


\begin{references} 
\bibitem{dgh}J. F. Donoghue, E. Golowich and B. R. Holstein,
{\it Dynamics of the standard model}, (Cambridge University Press, Cambridge,
1992). 
\bibitem{vmd}J. J. Sakurai, {\it Currents and mesons}, (University of
Chicago Press, Chicago, 1969). 
\bibitem{current}V. de Alfaro, S. Fubini, G. Furlan and C. Rossetti, 
{\it Currents in hadron physics}, (North Holland, Amsterdam, 1973).
\bibitem{gg}S. Gasiorowicz and D. A. Geffen, Rev.\ Mod.\ Phys.\ {\bf 41} (1969)
531. 
\bibitem{syracuse}\"O. Kaymakcalan, S. Rajeev and J. Schechter, Phys.\
Rev.\ {\bf D30} (1984) 594;\\
 H. Gomm, \"O. Kaymakcalan and J. Schechter, {\it ibid.}\ 2345. 
\bibitem{meissner}U.-G. Meissner, Phys.\ Rep.\ {\bf 161} (1988) 213. 
\bibitem{bky}M. Bando, T. Kugo and K. Yamawaki, Phys.\ Rep.\ {\bf 164} (1988) 
217. 
\bibitem{wein}S. Weinberg, Phys.\ Rev.\ {\bf 166} (1968) 1568.
\bibitem{ccwz}S. Coleman, J. Wess and B. Zumino, Phys.\ Rev.\ {\bf 177} (1969)
2239;\\ 
C. G. Callan, S. Coleman, J. Wess and B. Zumino, {\it ibid.}\ 2247.
\bibitem{ecker1}G. Ecker, J. Gasser, A. Pich and E. de Rafael, Nucl.\ Phys. 
\ {\bf B321} (1989) 425. 
\bibitem{ecker2}G. Ecker, H. Leutwyler, J. Gasser, A. Pich and E. de Rafael, 
Phys.\ Lett.\ {\bf B223} (1989) 425. 
\bibitem{enjl}J. Bijnens, Ch. Bruno and E. de Rafael, Nucl.\ Phys.\ {\bf 390} 
(1993) 501.
\bibitem{bijnens}J. Bijnens, NORDITA preprint 95/12NP (1995), hep-ph/9502393.
\bibitem{bgp}A. Bramon, A. Grau and G. Pancheri, Phys.\ Lett.\ {\bf 317} (1993)
190.
\bibitem{esk}S. I. Eidelman, Z. K. Silagadze and E. A Kuraev, 
Phys.\ Lett.\ {\bf B346} (1995) 186.
\bibitem{oconn}H. B. O'Connell, B. C. Pearce, A. W. Thomas and A. G. Williams,
Phys.\ Lett.\ {\bf B336} (1994) 1; {\bf B354} (1995) 14; University of
Adelaide preprint ADP-95-1/T168 (1995), hep-ph/9501251, to be published in
{\it Trends in Particle and Nuclear Physics};\\
H. B. O'Connell, A. G. Williams, M. Bracco and G. Krein,
University of Adelaide preprint ADP-95-49/T196 (1995), hep-ph/9510425.
\bibitem{gklsy}C. Gale and J. I. Kapusta, Nucl.\ Phys.\ {\bf B357} (1991) 65;\\
C. Song, Phys.\ Rev.\ {\bf D48} (1993) 1375;\\
S.-H. Lee, C. Song and H. Yabu, Phys.\ Lett.\ {\bf B341} (1995) 407;\\
C. Song, hep-ph/9501364.
\bibitem{sh}H. Shiomi and T. Hatsuda, Phys.\ Lett.\ {\bf B334} (1994) 281.
\bibitem{br}G. E. Brown and M. Rho, Phys.\ Lett.\ {\bf B338} (1994) 301;
hep-ph/9504250, to be published in Phys.\ Reports.
\bibitem{pisarski}R. D. Pisarski, Phys.\ Rev.\ {\bf D52} (1995) R3773;
hep-ph/9505257.
\bibitem{wpin}S. Weinberg, Phys.\ Rev.\ Lett.\ {\bf 17} (1966) 616.
\bibitem{tomo}Y. Tomozawa, Nuovo Cim.\ {\bf 46A} (1966) 707.
\bibitem{ksfr}K. Kawarabayashi and M. Suzuki, Phys.\ Rev.\ Lett. {\bf 16} 
(1966) 255;\\
Fayyazuddin and Riazuddin, Phys.\ Rev.\ {\bf 147} (1966) 1071. 
\bibitem{gl}J. Gasser and H. Leutwyler, Ann.\ Phys.\ {\bf 158} (1984) 142;
Nucl.\ Phys.\ {\bf B250} (1985) 465.
\bibitem{drv}J. F. Donoghue, C. Ramirez and G. Valencia, Phys.\ Rev.\ {\bf D39}
(1989) 1947.
\bibitem{kal}D. Kalafatis, Phys.\ Lett.\ {\bf B313} (1993) 115. 
\bibitem{kb}D. Kalafatis and M. C. Birse, Nucl.\ Phys.\ {\bf A584} (1995) 589. 
\bibitem{georgi}H. Georgi, Phys.\ Rev.\ Lett.\ {\bf 63} (1989) 1917; 
Nucl.\ Phys.\ {\bf 331} (1990) 311. 
\bibitem{bando2}M. Bando, T. Kugo and K. Yamawaki, Nucl.\ Phys.\ {\bf B259}
(1985) 493. 
\bibitem{bando3}M. Bando, T. Kugo and K. Yamawaki, Prog.\ Theor.\ Phys.\ {\bf 
73} (1985) 1541. 
\bibitem{bpr}Y. Brihaye, N. K. Pak and P. Rossi, Nucl.\ Phys.\ {\bf B254} 
(1985) 71.
\bibitem{pp}E. Pallante and R. Petronzio, Nucl.\ Phys.\ {\bf B396} (1993) 205.
\bibitem{bm}B. Borasoy and U.-G. Meissner, Universit\"at Bonn preprint
TK-95-31, hep-ph/9511320.
\bibitem{bp}J. Bijnens and E. Pallante, NORDITA preprint 95/63, hep-ph/9510338.
\bibitem{bando1}M. Bando, T. Kugo, S. Uehara, K. Yamawaki and T. Yanagida, 
Phys.\ Rev.\ Lett.\ {\bf 54} (1985) 1215.
\bibitem{pdg}Particle Data Group, Phys.\ Rev.\ {\bf D50} (1994) 1443.
\bibitem{bc}M. K. Banerjee and J. B. Cammerata, Phys.\ Rev.\ {\bf D17} (1978)
1125.
\bibitem{klz}N. M. Kroll, T. D. Lee and B. Zumino, Phys.\ Rev.\ {\bf 157} 
(1967) 1376.
\bibitem{skyrme}T. H. R. Skyrme, Proc.\ Roy.\ Soc.\ {\bf A260}
(1961) 127; Nucl.\ Phys.\ {\bf 31} (1962) 556. 
\bibitem{gw}J. Goldstone and F. Wilczek, Phys.\ Rev.\ Lett.\ {\bf 47} (1981) 
986.
\bibitem{witten}E. Witten, Nucl.\ Phys.\ {\bf B233} (1983) 422, 433.
\bibitem{wz}J. Wess and B. Zumino, Phys.\ Lett.\ {\bf B37} (1971) 95.
\bibitem{zb}I. Zahed and G. E. Brown, Phys.\ Reports {\bf 142} (1986) 1.
\bibitem{fro}M. Froissart, Phys.\ Rev.\ {\bf 123} (1961) 1053;\\
A. Martin, Phys.\ Phys.\ {\bf 129} (1963) 1432.
\bibitem{penn}M. R. Pennington and J. Portoles, Phys.\ Lett.\ {\bf B344} (1995)
399.
\bibitem{hn}K. Huber and H. Neufeld, Phys.\ Lett.\ {\bf B357} (1995) 221.
\bibitem{fnr79}E. G. Floratos, S. Narison and E. de Rafael, Nucl.\ Phys.\ {\bf
B155} (1979) 115.
\bibitem{hy}M. Harada and K. Yamawaki, Phys.\ Lett.\ {\bf B297} (1992) 151. 
\bibitem{hky}M. Harada, T. Kugo and K. Yamawaki, Phys.\ Rev.\ Lett.\ {\bf 71} 
(1991) 1299; Prog.\ Theor.\ Phys.\ {\bf 91} (1994) 801. 
\bibitem{plant}R. S. Plant, private communication.
\bibitem{schechter}J. Schechter, Phys.\ Rev.\ {\bf D34} (1986) 868. 
\bibitem{mz}U.-G. Meissner and I. Zahed, Z. Phys.\ {\bf A327} (1987) 5. 
\bibitem{yamaw}K. Yamawaki, Phys.\ Rev.\ {\bf D35} (1987) 412.
\bibitem{ks}\"O. Kaymakcalan and J. Schechter, Phys.\ Rev.\ {\bf D31} (1984)
1109. 
\bibitem{wrel}S. Weinberg, Phys.\ Rev.\ Lett.\ {\bf 18} (1967) 507.
\bibitem{kuraev}E. A. Kuraev and Z. K. Silagadze, Phys.\ Lett.\ {\bf B292}
(1992) 377;\\
E. L. Bratovskaya, E. A. Kuraev, Z. K. Silagadze and O. V. Teraev, 
Phys.\ Lett.\ {\bf B338} (1994) 471.
\bibitem{pb}R. S. Plant and M. C. Birse, Phys.\ Lett.\ {\bf B365} (1996) 292.
\bibitem{dual}P. K. Townsend, K. Pilch and P. van Nieuwenhuizen, 
Phys.\ Lett.\ {\bf B136} (1984) 38;\\
S. Deser and R. Jackiw, Phys.\ Lett.\ {\bf B139} (1984) 371.
\bibitem{malt}K. Maltman, Phys.\ Lett.\ {\bf B351} (1995) 507; hep-ph/9504237,
hep-ph/9504404. 
\bibitem{ghp}S. Gardner, C. J. Horowitz and J. Piekarewicz, 
Phys.\ Rev.\ Lett.\ {\bf 75} (1995) 2462; nucl-th/9508035.
\bibitem{mow}K. Maltman, H. B. O'Connell and A. G. Williams, University of
Adelaide preprint ADP-95-50/T197, hep-ph/9601309.
\bibitem{cm}T. D. Cohen and G. A. Miller, Phys.\ Rev.\ {\bf C52} (1995) 3428.
\bibitem{ijl}M. J. Iqbal, X. Jin and D. B. Leinweber, University of Washington
preprint UW-PP-DOE/ER/40427-19-N95, nucl-th/9504026.
\bibitem{malt2}K. Maltman, Phys.\ Lett.\ {\bf B362} (1995) 11.

\end{references}
\end{document}